\documentclass[11pt]{article}
\usepackage{graphics}
\usepackage{epsfig}
\usepackage{epstopdf}
\usepackage{amsmath}
\usepackage{array}
\usepackage[numbers,sort&compress]{natbib}
\usepackage{setspace}
\begin{document}
{\begin{center}	\bf{\LARGE On the feasibility of truncated Israel-Stewart model in the context of late acceleration } \end{center}}
\begin{center}
	Jerin Mohan N D and Titus K Mathew\\
	Department of Physics, Cochin University of Science and Technology\\
	Kochi - 22, India\\
{\it	jerinmohandk@cusat.ac.in, titus@cusat.ac.in}
\end{center}
\begin{abstract}
	A dissipative model of the Universe based on the causal relativistic truncated Israel-Stewart theory is analysed in the context of recent accelerated expansion of the Universe.
	The bulk viscosity and relaxation time are taken as $\xi=\alpha\rho^s$ and $\tau=\frac{\alpha}{\epsilon\gamma(2-\gamma)}\rho^{s-1}$ respectively. For $s=1/2,$ we found an analytical 
	solution for the Hubble parameter of the model.
	We have estimated the model parameters by treating $\gamma=1$ and $\gamma$ as a free parameter using the latest cosmological data. 
	The model 
	predicts 
	a prior decelerated phase and an end de Sitter phase as in the standard $\Lambda$CDM model. The dynamical system analysis shows that the prior decelerated epoch is an unstable 
	equilibrium, while the far future de Sitter epoch is stable. The age of the Universe obtained around $13.66$ Gyr, which is close to the 
	recent observations.  
	The second law of thermodynamics is found to be satisfied throughout the evolution in this model. The feasibility of the model has been checked by contrasting 
	with models based on the full Israel-Stewart and the Eckart viscous theories. The truncated viscous model appears more compatible with astronomical observations than the Eckart 
	and full causal viscous models.
\end{abstract}

\section{Introduction}
\label{sec:IntroductionTIS}
The bulk viscous phenomenon in cosmology has attracted considerable interest since this is the possible dissipative mechanism that can occur in a homogeneous and isotropic Universe. In such non-equilibrium thermodynamic processes, the traditionally used relativistic approach is due to Eckart \cite{Eckart}. However, the Eckart theory is suffering from the superluminal speed of the signals, 
which causes the violation of 
causality, and the equilibrium states are generally unstable 
(see \cite{Coley2,Israel1,Hiscock1,Avelino1} and references therein). The reason is that it accounts only up to first-order deviations from equilibrium. If one includes the higher-order deviations as well, then the problems may disappear. Motivated from this, a second-order relativistic theory of the dissipative process was formulated by Israel and Stewart (IS theory) \cite{Israel1,IsraelStewart,IsraelStewart2}. 
In this 
theory, the systems have a finite relaxation time $\tau$ to the equilibrium state in contrast to the 
Eckart theory, in which the system relaxes instantaneously. Due to this, the causality is restored in the Israel-Stewart theory, and also the equilibrium states become generally 
stable \cite{Hiscock,Hiscock1}.

An important point of interest is whether a negative bulk viscous pressure can cause an accelerated expansion in the 
FLRW Universe. The two accelerated expansion epochs in the evolution of the Universe are the early inflation and the late accelerated epoch. Several studies are there analysing the possibility 
of a bulk viscosity inflation. However, the results are converse with each other. Some authors argued in favour of the existence of such an inflationary solution  \cite{HiscockPRD1991,Prisco}, while others  negate such a possibility \cite{PavonCQG1991, ZakariPRD1993,Maartens}. There exist differences 
between the full and truncated theories in addressing the early inflation as discussed in \cite{Maartens}. In \cite{HiscockPRD1991,ZakariPRD1993}, the authors have concluded that the existence of an inflationary solution is a spurious effect that arises due to neglecting the non-linear terms in full IS theory i.e., due to the usage of truncated version. On the other hand, in the ref. \cite{ZimdahlTIS}, the authors have argued that both full and truncated versions will allow inflationary solutions under exceptional conditions even in situations far away from equilibrium. Qualitative and numerical studies of dissipative models by Romano and Pavon have shown that 
both truncated and full versions give an asymptotically stable de Sitter solution but an unstable FLRW solution \cite{RomanoandPavon}.

Regarding the possibility of bulk viscosity driven late acceleration, models based on both the Eckart theory \cite{Athira,Athira2} and the
full IS theory \cite{EPJC1,CQG2} are strongly supporting the possibility. But how far a truncated IS theory will support such a 
possibility is not studied adequately in the current literature. An analysis based on the truncated IS theory in the context of late acceleration is our main aim in this paper.

We will describe the relationship between the full IS theory and its truncated version in the next section. However, the following points may be noted from the outset. 
In non-equilibrium thermodynamics, usually called as the `extended irreversible thermodynamics'
\cite{ZimdahlTIS}, the generalised Gibbs equation, the evolution equation of thermodynamic variables involves thermodynamic quantities dissipative in nature. 
Then the corresponding causal evolution equation is approximately equivalent to a truncated theory expressed in terms of the equilibrium variables \cite{GarielPRD1994}.
For analysis in situations like expanding Universe, such a truncated version is equivalent to the full causal theory 
due to Israel and Stewart \cite{Pavon1982,ZakariPRD1993,Maartens}. The essential fact is that the truncated theory retains stability and causality conditions. 
Therefore, it is reasonable to consider the truncated Israel-Stewart theory, as an independent theory to describe the
bulk viscous pressure evolution \cite{Piattella}. The truncated Israel-Stewart formalism is also called the Maxwell-Cattaneo theory.

Even though the $\Lambda$CDM model explains the current acceleration in the expansion of the Universe \cite{Sami1}, it faces the following issues. 
First, it admits a fictitious component called dark energy to realise the late acceleration, but its nature is still a mystery. The model admits 
the cosmological constant as the dark energy; nevertheless, its estimated value is very tiny compared to that predicted by the quantum field theory \cite{Sami1}. 
Secondly, it doesn't give any valid explanation for the equality of the densities of the evolving dark matter and the cosmological constant during the 
current epoch \cite{Sami1}. Bulk viscosity driven acceleration is an alternative approach in finding reasonable solutions to these problems. 
A suitable mechanism for the origin of bulk viscosity in an isotropic and homogeneous Universe is still not clear. However, alternative 
speculations were made in the literature. Some authors have shown that different cooling rates of the components 
of the cosmic medium can produce 
bulk viscosity \cite{Weinberg2, Schweizer, Udey, Zimdahl3} or decay of dark matter particles into relativistic particles \cite{JRWilson} can be 
a reason for bulk viscosity.
Another proposal is that bulk viscosity of the cosmic fluid maybe 
the result of non-conserving particle interactions \cite{Murphy, Turok, Zimdahl4}. In \cite{RomanoandPavon}, Romano and Pavon have speculated that dissipative effects, 
which generate viscous pressure, can arise due to the non-equilibrium effects result from the quark-gluon interaction \cite{Thoma}. 
Interaction between the different components of dark matter may generate bulk and shear stresses \cite{PavonZimdahl}. 
In \cite{Gimenes},
the viscosity through the introduction of negative pressure due to matter creation is analysed, using which the authors have described accelerating Universe 
without considering the conventional dark energy component.

For explaining the recent acceleration of the Universe \cite{Reiss,Perlmutter}, a model of bulk viscous matter dominated Universe based on the
Eckart formalism is discussed in ref. \cite{Avelino1,Avelino2,Athira,Athira2}. 
In particular, the authors in \cite{Athira2} have shown that the bulk viscous matter dominated model can predict a transition into a late accelerated epoch, which asymptotically tends to a stable de Sitter phase.

Owing to the non-causal nature of the Eckart formalism, the bulk viscosity driven late acceleration has been analysed using the full causal Israel-Stewart theory \cite{EPJC1,CQG2,Cruz2,Cruz3,Dmitry,Dmitry2}. For instance, the model described in \cite{EPJC1} predicts the transition into a late accelerated phase, where the Universe has a quintessence nature in the late phase evolution. However, the model failed to predict a pure de Sitter phase as the end stage. The dynamical system anlaysis of this model presented in \cite{CQG2} shows that the model can admit an asymptotically stable accelerated phase.

Some other works of important relevance to the viscous nature of the dark sectors in resolving some problems related to the recent observations are discussed in \cite{Brevik5,Arvind,Arvind2,anand1}. 
In \cite{anand1}, the authors have shown that the $\sigma_8-\Omega_{m}$ tension (where $\sigma_8$ is the r.m.s. fluctuations of perturbations at 
$8{h}^{-1}$Mpc scale) and $H_0-$ $\Omega_m$ tension faced in the analysis of the Planck CMB parameters using the standard $\Lambda$CDM model can alleviate if one assumes a small amount of viscosity in the dark matter sectors. To mention another one, the amplitude of the absorption signal of $21$ cm line at the redshift 
$z\sim 17$ announced in the EDGES (Experiment to Detect the Global EoR Signature) observation is larger than the prediction from the standard 
cosmological model \cite{Monsalve}. In \cite{Arvind}, the authors impart viscous nature to the dark matter for explaining this anomaly in the 
EDGES observation. Recently with reference to the gravitational wave observations \cite{Abbott1,Abbott2}, the impacts of propagation 
of the gravitational wave 
signal in the cosmic fluid with viscosity, during the early and late epoch of the Universe, is reported by Brevik and Nojiri in \cite{Brevik5} and 
they have constrained the viscosity of the cosmic fluid using the observational data. 
In the review \cite{Brevikrev1}, the influence of viscous effects in the expansion profile of the Universe are given in detail.

As mentioned earlier, our aim in this paper is to make a detailed analysis of the bulk viscosity driven late acceleration by using the truncated version of IS theory and hence to see whether there arise any difference in total effect as compared to the usage of the full IS theory as discussed in \cite{EPJC1,CQG2}. The reason for this interest is due to the difference in the conclusions regarding the bulk viscous inflation between the full and truncated theories \cite{Maartens}. The authors 
in \cite{Piattella} have analysed the evolution 
of gravitational potential by assuming an ansatz on the pressure of viscous fluid using the truncated 
IS theory. In their work, they have solved the truncated form of the viscous pressure evolution equation to obtain 
the relaxation time of the viscous fluid from a non-equilibrium state to an equilibrium state and then proceed with 
the consideration of the evolution of potential \cite{Piattella}. They have shown that the truncated version is compatible with 
the standard $\Lambda$CDM model in accounting for the CMB power spectrum. We will mainly concentrate on the Hubble parameter evolution and the status of other relevant cosmological parameters. For obtaining an analytical solution for the Hubble parameter, we have assumed the
relation between bulk viscosity $\xi$ and the energy density $\rho$ of the form $\xi=\alpha \rho^s,$ where $\alpha$ is a constant \cite{Maartens,Belinskii1976,Belinskii1977}. For the parameter $s,$ we have considered the standard cases, $s=1/2$ and $s\neq1/2.$

This paper is organized as follows. In section \ref{sec:Bulk viscous matter dominated Universe}, the analytical solution of the Hubble parameter and the analysis of cosmological parameters are given. The status of near equilibrium condition is given in section \ref{sec:nearequilibriumconditionTIS}. In section \ref{sec:PhasespaceanalysisTIS}, we present the phase space analysis of the model. Section \ref{sec:ThermodynamicanalysisTIS} deals with thermodynamic analysis of the model.
Section \ref{sec:EstimationofmodelparametersTIS} presents the model parameters estimation, and the conclusions are given in section \ref{sec:ConclusionsTIS}.

\section{Evolution of the viscous Universe with truncated IS theory}
\label{sec:Bulk viscous matter dominated Universe}
We consider a flat Universe with the bulk viscous matter as the dominant cosmic component, satisfying the Friedmann equations, 
\begin{equation} \label{eqn:F1}
H^{2}=\frac{\rho}{3},
\end{equation}
\begin{equation}
\label{eqn:F2}
\dot{H}=-H^2-\frac{1}{6}\left(\rho+3P_{eff}\right),
\end{equation}
where $H=\frac{\dot{a}}{a}$ is the Hubble parameter, $ \rho $ is the matter density, $ P_{eff} $ is the effective pressure, 
$a$ is the scale factor, an over dot represents the derivative with respect to cosmic time $t,$ and we have taken $c= 8\pi G=1.$ The conservation equation for the viscous fluid is
\begin{equation} \label{eqn:con1}
\dot{\rho}+3H(\rho+P_{eff})=0.
\end{equation}
The effective pressure is given as,
\begin{equation}
\label{eqn:effectivep}
P_{eff}=p+\Pi,
\end{equation}
where $ p $ is the normal pressure, given by $ p=(\gamma-1)\rho $, $ \gamma $ is the barotropic index and $ \Pi $ is the bulk viscous pressure. In the full causal Israel-Stewart theory, the bulk viscous pressure satisfies the dynamical evolution equation,
\begin{equation}\label{eqn:IS1}
\tau\dot\Pi+\Pi=-3\xi H-\frac{1}{2}\tau\Pi\left(3H+\frac{\dot\tau}{\tau}-\frac{\dot\xi}{\xi}-
\frac{\dot{T}}{T}\right),
\end{equation}
where $\tau$, $\xi$ and $T$ are the relaxation time, bulk viscosity and temperature of the viscous fluid respectively and are generally functions of the density of fluid \cite{Maartens}. As $\tau \to 0,$ this evolution equation will reduces to the Eckart equation, $\Pi = -3\xi H.$

The truncated version is an approximation of the full IS theory, resulting when the bracketed terms on the RHS of (\ref{eqn:IS1}) are assumed to be negligible in comparison with the viscosity term $-3H\xi.$ Consequently, the evolution of viscous pressure will be represented as, 
\begin{equation}
\label{eqn:TIS}
\tau\dot\Pi+\Pi=-3\xi H.
\end{equation}
This approximation is undoubtedly correct under situations when the system is very close to the equilibrium state \cite{ZimdahlTIS}. However the full IS theory and truncated form are exactly identical to each other if $\left(3H+\frac{\dot\tau}{\tau}-\frac{\dot\xi}{\xi}- \frac{\dot{T}}{T}\right) =0,$ which is satisfied only in some exceptional cases (for details of this see \cite{ZimdahlTIS}). Even otherwise, the truncated version will suitably hold if the viscous pressure satisfies the condition $|\Pi| \ll \rho$ \cite{ZimdahlTIS}. The important advantage of the truncated form is that (i) it does not contain the complicated non-linear terms as in (\ref{eqn:IS1}), (ii) like the full theory, it also predicts a finite speed for the viscous pulses hence safeguard the principle of causality \cite{ZakariPRD1993,MaartensTIS}.

Following Belinskii et al. \cite{Belinskii1976,Belinskii1977}, we choose the coefficient of viscosity as,
\begin{equation}
\label{eqn:xi}
\xi=\alpha\rho^{s}.
\end{equation}
and the relaxation time as $\tau=\xi/\rho$, 
here $\alpha$ and $s$ are constant parameters satisfying the
conditions $\alpha\geq0$ and $s\geq0.$ However from the causality and stability
conditions of the IS theory, a general expression for $\tau$ can be derived \cite{MaartensTIS},
\begin{equation}
\label{eqn:viscosityrelaxationtimeTIS}
\frac{\xi}{(\rho+p)\tau}=c_b^2,
\end{equation} 
where $c_b$ represents the speed of bulk viscous perturbations. The dissipative
speed of sound $v$ can be expressed as, $v^2=c_s^2+c_b^2,$ where $c_s$ is the adiabatic sound speed and is given as $c_s^2=\frac{\partial{p}}{\partial{\rho}},$ since the normal pressure is $p=(\gamma-1)\rho$, then the adiabatic sound speed is obtained as $c_s^2= \gamma-1$ for a barotropic fluid.
Hence to satisfy the causality condition, $v^2 \leq 1,$ the speed of bulk viscous perturbations takes the form $c_b^2=\epsilon(2-\gamma)$ with $0<\epsilon\leq1$ \cite{MaartensTIS}. Using (\ref{eqn:xi}) and causality conditions of the bulk viscous perturbations, (\ref{eqn:viscosityrelaxationtimeTIS}) can be modified as,
\begin{equation}
\label{eqn:viscosityrelaxationtimeTIS2}
\frac{\alpha\rho^s}{\left[\rho+(\gamma-1)\rho\right]\tau}=\epsilon(2-\gamma),
\end{equation}
Now, from (\ref{eqn:viscosityrelaxationtimeTIS2}) the relaxation time $\tau$ becomes \cite{MaartensTIS,Piattella}
\begin{equation}
\label{eqn:tau}
\tau=\frac{\alpha}{\epsilon \gamma(2-\gamma)}\rho^{s-1}.
\end{equation}
Friedmann equation (\ref{eqn:F1}) can then combined with (\ref{eqn:con1}) and (\ref{eqn:effectivep}), to express the bulk viscous pressure $\Pi$ as
\begin{equation}
\label{eqn:pi}
\Pi=-\left[2\dot{H}+3H^2+(\gamma-1)\rho\right],
\end{equation}
and its time derivative is,
\begin{equation}
\label{eqn:dotpi}
\dot{\Pi}=-\left[2\ddot{H}+6H\dot{H}+(\gamma-1)\dot{\rho}\right].
\end{equation}
Using (\ref{eqn:xi}), (\ref{eqn:tau}), (\ref{eqn:pi}) and (\ref{eqn:dotpi}), the bulk viscous pressure evolution (\ref{eqn:TIS}) can be modified as,
\begin{eqnarray}
\label{eqn:Hddot1}
\ddot{H}+3\gamma H \dot{H}+\frac{3^{1-s}\epsilon\gamma(2-\gamma)}{\alpha}H^{2-2s}\dot{H}+\frac{3^{2-s}\epsilon\gamma^2 (2-\gamma)}{2\alpha}H^{4-2s} \nonumber\\
-\frac{9}{2}\epsilon\gamma(2-\gamma)H^3 =0.
\end{eqnarray}
We are interested in the late Universe in which matter is non-relativistic, for which $\gamma=1$ (then velocity of bulk viscous perturbation, $c_b^2=\epsilon(2-\gamma)=\epsilon).$ Numerically $\epsilon$ is very small and an analysis based on the truncated IS theory shows that $10^{-11}\ll\epsilon\leq 10^{-8}$ \cite{Piattella}, and we also took $ s=1/2$ \cite{ChimentoJacubi}, which according to (\ref{eqn:xi}) 
implies that the bulk viscosity is directly proportional to the Hubble parameter. Now  (\ref{eqn:Hddot1}) becomes,
\begin{equation}
\label{eqn:HdifferentialequationTIS}
\ddot{H}+b_1 H\dot{H}+b_2 H^3=0,
\end{equation}
where
\begin{equation}
b_1=3\gamma\left[\frac{\epsilon(2-\gamma)}{\sqrt{3}\alpha}+1\right], \, b_2=\frac{9\epsilon\gamma(2-\gamma)}{2}\left(\frac{\gamma}{\sqrt{3}\alpha}-1\right).
\end{equation}
For the calculation purpose we change the variable from cosmic time $t$ to $ x = \ln a $, so that differential equation (\ref{eqn:HdifferentialequationTIS}) becomes
\begin{equation}
\label{eqn:DE}
\frac{d^2H}{dx^2}+\frac{1}{H}\left(\frac{dH}{dx}\right)^2+b_1\frac{dH}{dx}+b_2H=0.
\end{equation}
We solve this equation and obtained the Hubble parameter as
\begin{equation}
\label{eqn:HTIS}
H=H_0C_{1}a^{-m_1}\sqrt{\cosh\left[m_2(\log a -C_2)\right]},
\end{equation}
where the constants are given as,
\begin{equation}
m_1=\frac{b_1}{4},\,\,\,\,\,m_2=\frac{1}{2}\sqrt{b_1^2-8b_2},
\end{equation}
\begin{equation}
C_1=\frac{1}{{\sqrt{\cosh(m_2 C_2)}}},\,\,\,\,\,
C_2=\frac{1}{m_2}\tanh^{-1}\left[\frac{3(\tilde{\Pi}_0+1)-2 m_1}{m_2}\right],
\end{equation}
where $\tilde{\Pi}_{0}=\frac{\Pi}{3H_0^2}$ is a dimensionless bulk viscous pressure parameter. A little bit of calculations will show that $H \propto H_0 (1+ \textrm{const.} \, a^{-3})^{1/2},$ for an extremely small $\epsilon$ value. Then, as $a\rightarrow\infty,$ $H\sim $ constant, indicates the possibility of an end de Sitter phase. 
In the prior phase corresponding to an equivalent limit $a\rightarrow 0,$ 
the Hubble parameter behaves as $H\sim a^{-3/2},$ which implies a prior decelerated 
matter dominated epoch. From an earlier analysis of the full IS viscous model $H\sim a^{-2.8}$ as $a\rightarrow0$ and $H\sim a^{-0.2}$ as $a\rightarrow\infty$ depicts the far future quintessence behaviour \cite{CQG2}. The Hubble parameter in the Eckart viscous model becomes $H\sim a^{-3.4}$ as $a\rightarrow 0$ and $H\sim$ constant as $a\rightarrow\infty$ implies the far future de Sitter epoch. The behaviour of the fluid in both full IS, and Eckart viscous model is different from the evolution of ordinary matter. The evolution of the Hubble parameter characterises the difference in these viscous models in addressing the late phase evolution.

The present truncated viscous model assumes a single cosmic component, then the density parameter becomes $\Omega_{\textrm{{\small total}}}\sim \Omega_{\textrm{{\small dark matter}}},$ therefore from the Hubble parameter  (\ref{eqn:HTIS}) the matter density parameter $\Omega_{m}$ is obtained as 
\begin{equation}
\Omega_{m}=\frac{\rho_m}{\rho_{critical}}=\frac{H^2}{H_0^2}=C_1^2a^{-2 m_1}{\cosh\left[m_2(\log a -C_2)\right]},
\end{equation}
in the present time, $a=1,$ $\Omega_m$ for the best estimated values of the model parameters (presented in a later section) is obtained as,
\begin{equation}
\Omega_{m0}=C_1^2{\cosh (m_2C_2)}=1.
\end{equation}
\subsection{Scale factor and age of the Universe} 
\begin{figure}
	\centering
	\includegraphics[scale=0.5]{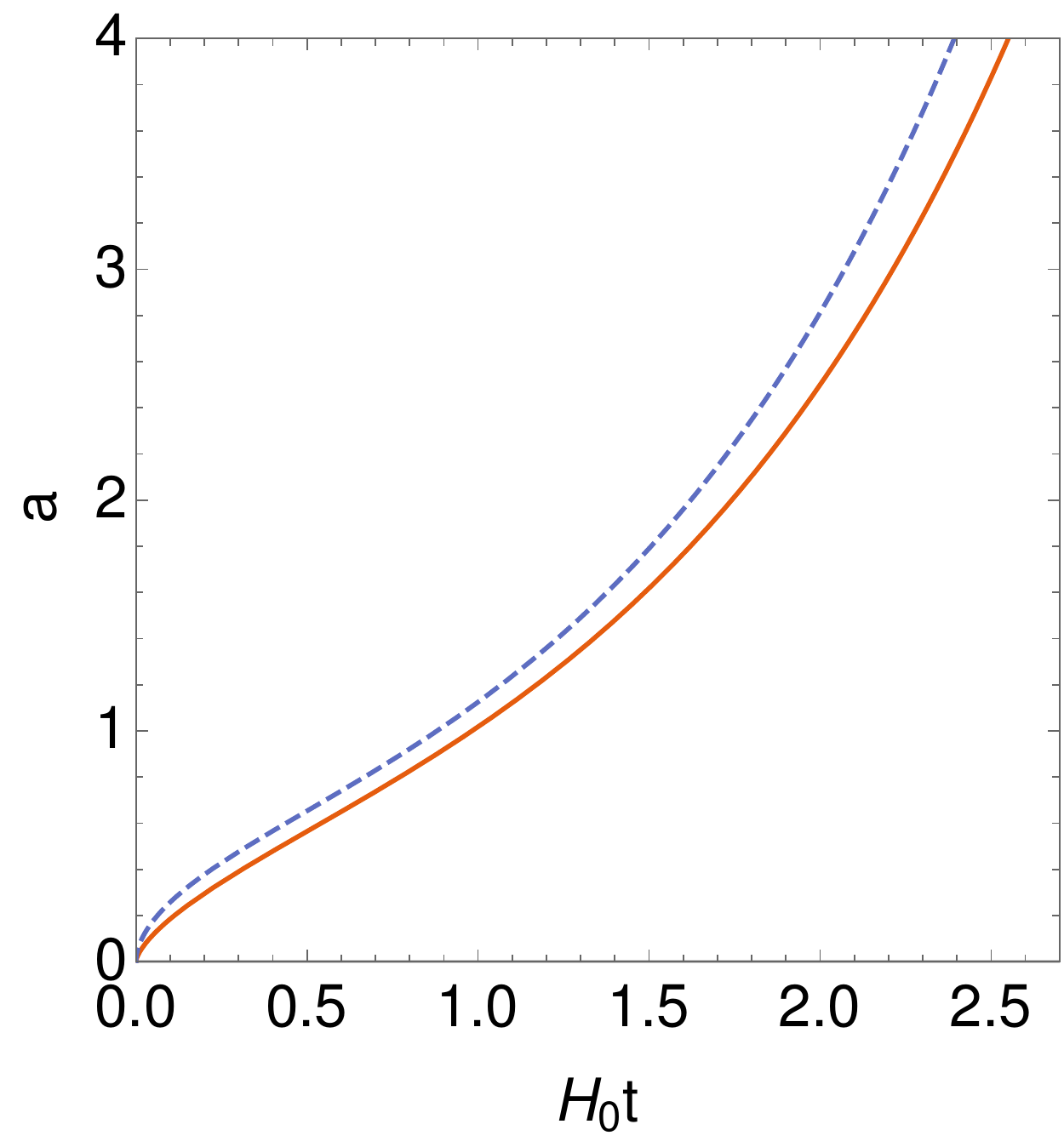}
	\caption{The evolution of scale factor with $H_0$t for the best estimated value of the model parameters for $\gamma=1$ (continuous line) and $\gamma=1.23$ (dashed line).}
	\label{fig:at}
\end{figure}

The Hubble parameter (\ref{eqn:HTIS}) can be expressed in a more convenient form as,
\begin{equation}
\label{eqn:HTISmodified}
H=H_0  \left(\tilde{C_1}a^{\tilde{-m_1}} + \tilde{C_2}a^{\tilde{-m_2}}\right)^{1/2},
\end{equation}
where 
\begin{equation}\label{eqn:newmTIS}
\tilde{m_1}=2m_1-m_2,\quad
\tilde{m_2}=2m_1+m_2.
\end{equation}\begin{equation}\label{eqn:newCTIS}
\tilde{C_1}=\frac{C_1^2e^{-m_2 C_2}}{2},\quad
\tilde{C_2}=\frac{C_1^2e^{m_2 C_2}}{2}.
\end{equation}
Integration of (\ref{eqn:HTISmodified}) gives,
\begin{eqnarray}
\label{eqn:aTIS}
\frac{2\left(1+{\frac{\tilde{C_1}a^{\tilde{m_2}-\tilde{m_1}}}{\tilde{C_2}}}\right)^{1/2}{_2F_1}\left(\frac{1}{2},-\frac{\tilde{m_2}}{2(\tilde{m_1}-\tilde{m_2})};1-\frac{\tilde{m_2}}{2(\tilde{m_1}-\tilde{m_2})};-\frac{\tilde{C_1}a^{\tilde{m_2}-\tilde{m_1}}}{\tilde{C_2}}\right)}{ \tilde{m_2}\left(\tilde{C_1}a^{\tilde{-m_1}} + \tilde{C_2}a^{\tilde{-m_2}}\right)^{1/2}} \nonumber\\ =H_0 (t_0-t_B),
\end{eqnarray} 
where ${_2F_1}$ represents a hypergeometric function, $t_0$ is the present time and $t_B$ is the time corresponding to the big-bang. To obtain the behaviour of scale factor, we 
made a parametric plot of (\ref{eqn:aTIS}), shown in figure \ref{fig:at}, for the best estimated values of model parameters. The figure shows a transition from the prior decelerated epoch to a later 
exponential evolution representing the end de Sitter epoch. 
The transition redshift $z_T,$ corresponding to the switching from the decelerated to the accelerated expansion can be obtained by equating the derivative, $\frac{d\dot a}{da}$ to zero. From  (\ref{eqn:HTISmodified}) we have 
\begin{equation}
\label{eqn:adotbyaTIS}
\frac{d\dot{a}}{da}=H_0\frac{\tilde{C_1}(2-\tilde{m_1})a^{1-\tilde{m_1}}+\tilde{C_2}(2-\tilde{m_2})a^{1-\tilde{m_2}}}{2\left(\tilde{C_1}a^{2-\tilde{m_1}}+\tilde{C_2}a^{2-\tilde{m_2}}\right)^{1/2}}=0.
\end{equation}
and is satisfied with $a=a_T$ and $z=z_T$ the transition scale factor $a_T$ and transition redshift $z_T$ respectively, and are found to be, 
\begin{equation}\label{eqn:zTTIS}
a_T=\left[-\frac{\tilde{C_2}(2-\tilde{m_2})}{\tilde{C_1}(2-\tilde{m_1})}\right]^{\frac{1}{\tilde{m_2}-\tilde{m_1}}},\quad z_T=\left[-\frac{\tilde{C_2}(2-\tilde{m_2})}{\tilde{C_1}(2-\tilde{m_1})}\right]^{-\frac{1}{\tilde{m_2}-\tilde{m_1}}}-1.
\end{equation}
For the best fit model parameter values, we get $z_T=0.73,$ which is in concordance with the WMAP observation 
\cite{UAlam}.

Unlike in the above calculation, we can obtain the age of the Universe by integrating  (\ref{eqn:HTISmodified}) within the limit from $a=0$ to $a=1.$ 
Then on substituting the best estimated values of the model parameters, we get the age as
\begin{equation}\label{eqn:AgeTIS}
t_0-t_B=\frac{98.28\times10^{-2}}{H_0}\sim 13.66\,\textrm{ Gyr}.
\end{equation}
This is very near to the age obtained from observations of the oldest globular clusters \cite{Carretta} and also 
matching with Planck observations, around $13.79 \textrm{ Gyr}$  \cite{AghaninPlanck2018}. 
For $\gamma\neq 1,$ the age of the Universe predicted from this model is around $12.25\,\textrm{Gyr}.$ The important point to be noted at 
this juncture is that the age predicted 
by the Eckart approach is around $10.9 \textrm{ Gyr}$ \cite{Athira} and that based on full Israel-Stewart 
formalism is around $9.72 \textrm{ Gyr}$ \cite{EPJC1}. These indicate the better performance of the truncated model.

\subsection{Evolution of the deceleration parameter}
We obtained deceleration parameter $q,$ characterising the rate of change of speed of expansion of the Universe. It is basically defined as
$q= -1-\dot H/H^2.$ Substituting the Hubble parameter and its time rate, we get
\begin{equation}
\label{eqn:qTIS}
q=-1+m_1-\frac{m_2}{2}\tanh\left[m_2(\log a-C_2)\right].
\end{equation}
As an approximation, for small value of $\epsilon,$ it can be obtained as, $q\sim-1+\frac{3}{4}-\frac{3}{4}\left( \frac{a^{3/2}-a^{-3/2}}{a^{3/2}+a^{-3/2}}\right).$ 
It is then clear that for large values of $a$ at which $a^{3/2}\gg a^{-3/2},$ the deceleration parameter attained a de Sitter type value, $q\sim -1.$ 
Whereas in the early expansion phase as $a\rightarrow 0,$ it follows $a^{-3/2}\gg a^{3/2},$ then $q\sim 1/2, $ which represents the early decelerated epoch. 
From (\ref{eqn:qTIS}), the current value $q,$ for the best fit values of the parameters 
(from the Pantheon sample) 
is obtained as $q_0\sim-0.58$ for $\gamma=1$ and $q_0\sim -0.64$ for $\gamma=1.23.$ This is 
matching with the WMAP value, $q_0=-0.60$ \cite{UAlam}. 

The evolution of the $q$ parameter with redshift for the best fit values is plotted in figure \ref{fig:qt}. The figure shows that the Universe makes a transition into the accelerating epoch from a prior decelerated epoch at the redshift around $z_T\sim0.73$ for the choice $\gamma=1$ and finally attaining the pure de Sitter epoch. The transition redshift for the choice $\gamma\neq1$ is obtained around $z_T\sim 0.54.$ It should be noted that in the bulk viscous model based on full causal formalism, it has been shown that \cite{CQG2}, the $q$ parameter stabilises around $q\sim-0.82,$ and hence never approach a pure de Sitter phase in the asymptotic limit. However, the Eckart viscous model asymptotically approaches a de Sitter phase \cite{Athira}.
\subsection{Evolution of the equation of state parameter}
Yet another parameter of interest is the equation of state of the cosmic fluid. The equation of state parameter can be obtained using the standard formula \cite{Praseetha}, 
\begin{equation}
\omega=-1-\frac{1}{3}\frac{d \ln h^{2}}{dx},
\end{equation}
where $h=\frac{H}{H_{0}}.$ 
Using the Hubble parameter in (\ref{eqn:HTIS}), the $\omega$ takes the form,
\begin{equation}
\omega=-1+\frac{1}{3}\left\{2m_1 - m_2 \tanh\left[m_2(\log a-C_2)\right]\right\}.
\label{eqn:EoSTIS}
\end{equation}
The above equation can be approximated as, $\omega\sim-1+\frac{1}{2}\left[1-\left(\frac{a^{3/2}-a^{-3/2}}{a^{3/2}+a^{-3/2}}\right)\right],$ for a small value of $\epsilon.$ This can 
readily follow that,
as $a\rightarrow\infty,$ $\omega \to -1$ and as $a\rightarrow 0,$ $\omega \to 0,$ which implies that the Universe will eventually attain a de Sitter phase from the decelerated expansion phase. From (\ref{eqn:EoSTIS}), the present value of $\omega$ for the best estimated values of the model parameters is obtained as $
\omega_0\sim-0.72$ for $\gamma=1$ and $\omega_0=-0.76$ for $\gamma=1.23.$ In this case, the current value is comparatively large compared to the value from the combined data set WMAP+BAO+$H_0$ +SN, about $\omega_0\sim-0.93$ \cite{Komatsu,Chimanto}.

\begin{figure}
	\centering
	\includegraphics[scale=0.56]{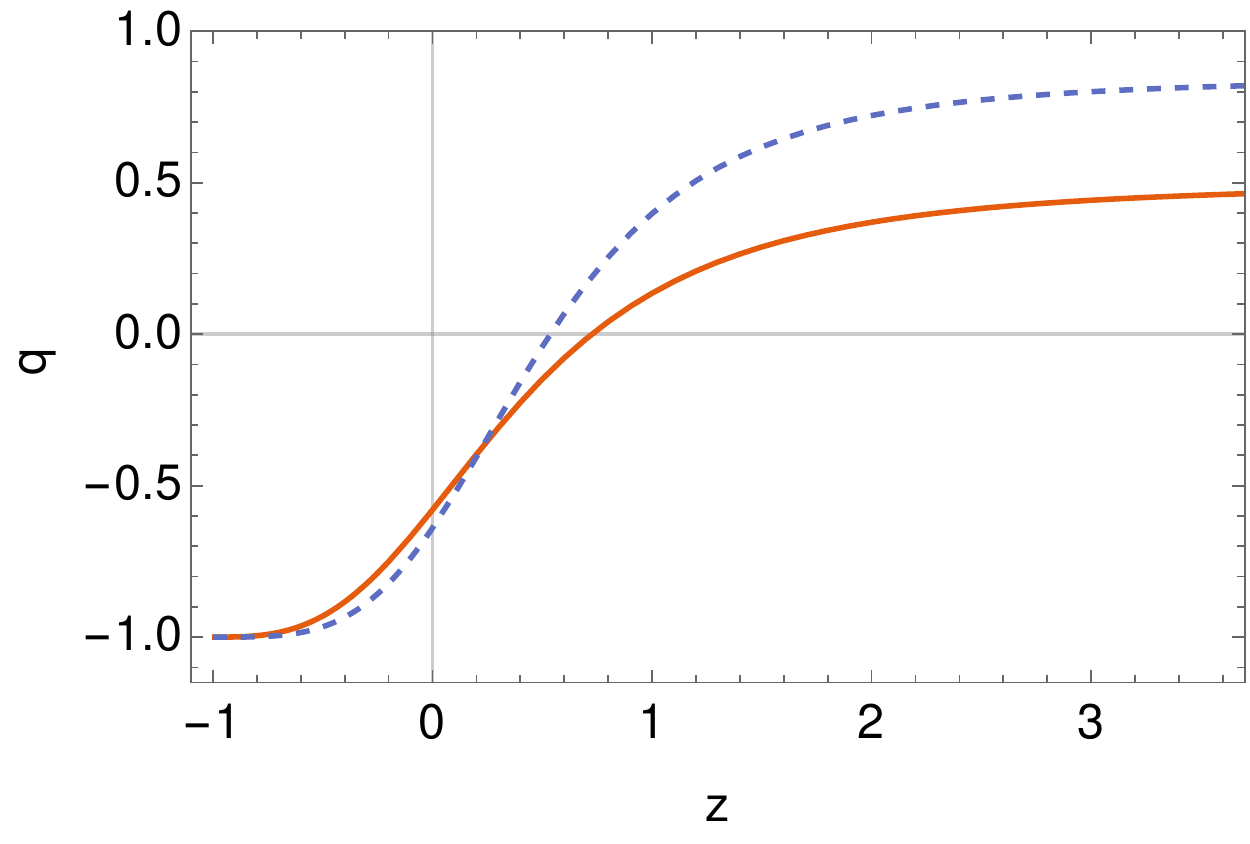}
	\hfill
	\includegraphics[scale=0.56]{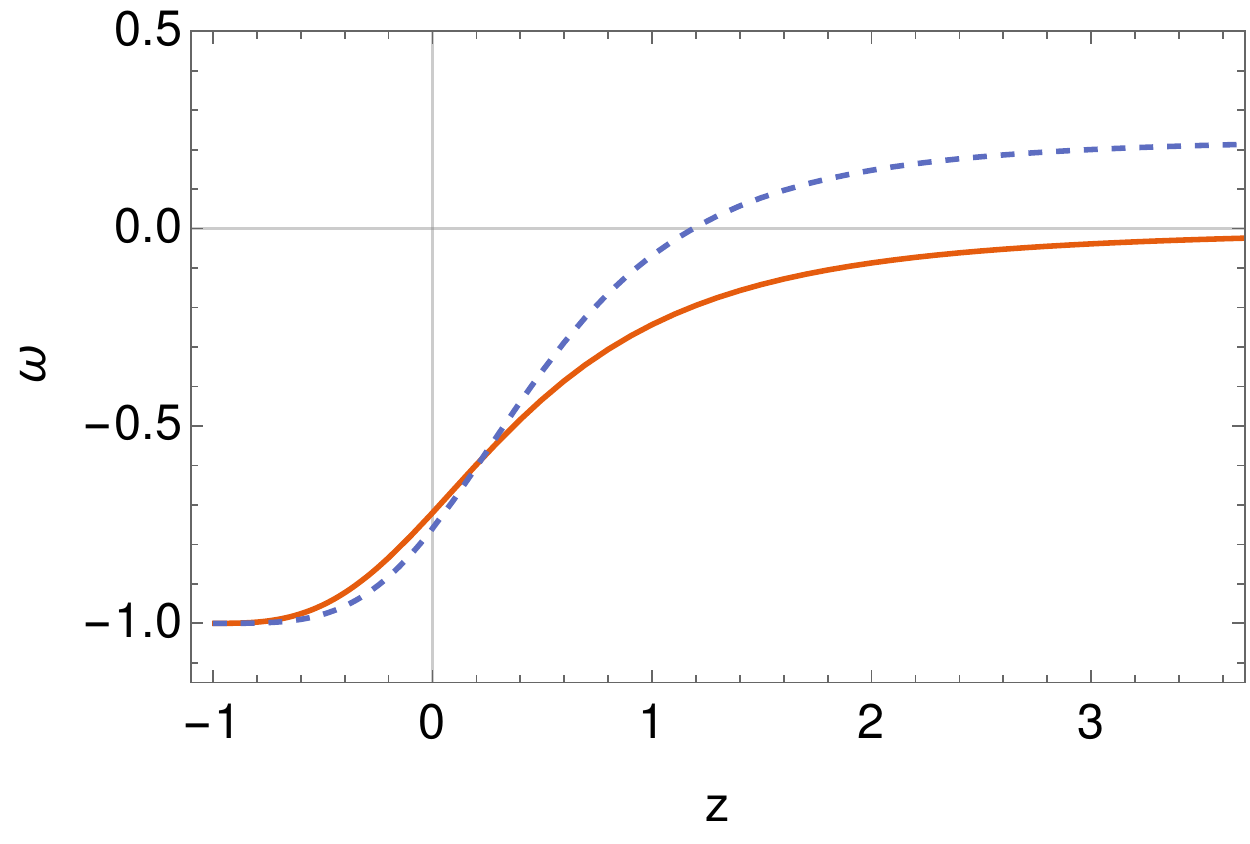}
	\caption{The evolution of deceleration parameter q, left-hand panel, and equation of state parameter $\omega,$ right-hand panel, with redshift for the best estimated value of model parameters for $\gamma=1$ (continuous line) and $\gamma=1.23$  (dashed line).}
	\label{fig:qt}
\end{figure}

The evolution of $\omega$ versus redshift for the best estimated values of the model parameter is shown in figure \ref{fig:qt}. The figure indicates an evolution of the viscous Universe from a prior decelerated expansion (at which $\omega \sim 0$) to the final de Sitter epoch 
with $\omega\sim-1.$
However, a model using the full causal IS theory \cite{EPJC1,CQG2} exhibits the quintessence nature in the far future evolution with $\omega\sim -0.88$ and stiff fluid character in the early decelerated phase of the Universe, $\omega\sim 0.88$. The early decelerated epoch in the Eckart viscous model also exhibits the stiff fluid nature $\omega\sim1.3,$ even though this model attains a far future de Sitter phase. Therefore, the model based on the truncated IS theory gives evolution similar to the standard cosmological model.
\section{Near equilibrium condition}
\label{sec:nearequilibriumconditionTIS}
\begin{figure}
	\centering
	\includegraphics[scale=0.67]{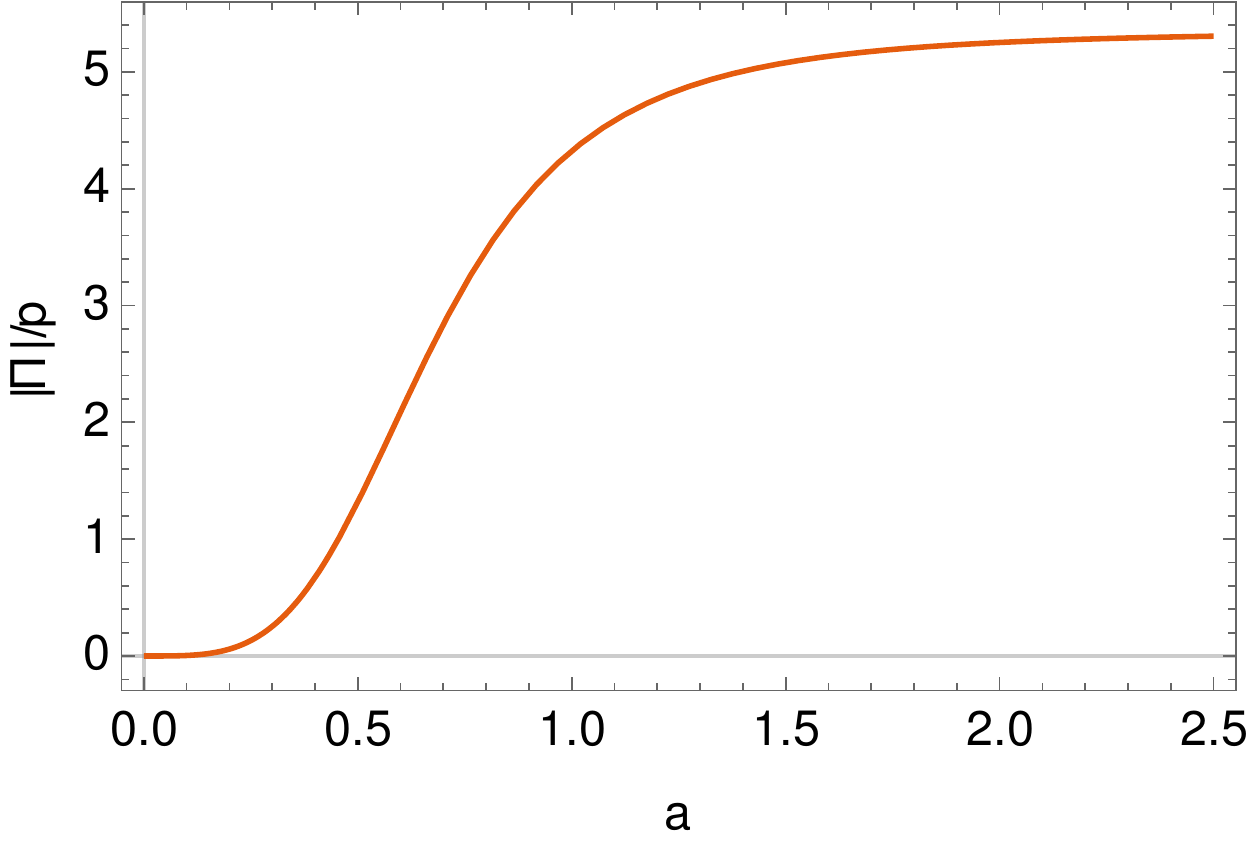}
	\caption{The evolution of $|\Pi|/p$ with scale factor for the best estimated values of the model parameters.}
	\label{plot:PvispTIS}
\end{figure}

The Israel-Stewart theory is formulated under the basic assumption that the thermodynamic state of the fluid is near to 
equilibrium i.e., $|\Pi|\ll p.$ From the second Friedmann equation (\ref{eqn:F2}), 
the condition 
accelerated expansion, $\ddot{a}>0,$ implies 
$-\Pi>\frac{\rho}{3}+p.$  This indicates that the fluid is away from equilibrium. Hence
it is reasonable to assume that 
the causal thermodynamics holds in the model beyond
the near-equilibrium regime.
It also implies that the viscous pressure is greater than the normal equilibrium pressure $p.$ In \cite{Maartens}, Maartens has suggested that causal Israel-Stewart 
theories hold beyond the near equilibrium regime since the fluid is far from equilibrium 
during the acceleration expansion. In a recent study \cite{Bemfica}, 
Bemfica et al. have proposed that Israel-Stewart like theories of fluid dynamics to be causal far away from equilibrium.
Furthermore, we analysed the evolution of $\Pi/p$ in the model corresponding to $\gamma\neq1$ and is given in figure \ref{plot:PvispTIS}. The plot shows that 
the near equilibrium condition is being violated, especially during the late stages of the Universe. 
As mentioned earlier, an accelerated expansion necessitates the violation of this particular 
condition.
\section{Phase space analysis}
\label{sec:PhasespaceanalysisTIS}
Understanding the global behaviour of the cosmological model can be obtained from the dynamical system study of the model. A cosmological model can be expressed as a system of autonomous differential equations by choosing suitable dynamic variables. Analysis of critical (or equilibrium) points obtained from the dynamical equations and the evolution of phase space trajectories give a general character of the cosmological model \cite{Ellis}. The critical points in the phase space can be classified as unstable (past attractor), stable (future attractor), saddle point, etc. to extract the evolutionary properties of respective epochs.

To obtain the phase space dynamics of the present model of the Universe, we use new dimensionless variables, $\Omega,$ the density parameter, $\tilde{\Pi},$ the bulk viscous pressure, and $\tilde{\tau}$ a new time. These are defined as, 
\begin{equation}
\label{eqn:dimensionlessphasespacevariable}
\Omega=\frac{\rho}{3H^2},
\,\quad
\tilde{\Pi}=\frac{\Pi}{3H^2},
\,\quad \textrm{and}\quad
H(t)dt=d\tilde{\tau}.
\end{equation}
Using the above definitions, (\ref{eqn:F2}), (\ref{eqn:con1}), and (\ref{eqn:TIS}) can be written as the dynamical system of equations,
\begin{equation}
\label{eqn:omega1}
\Omega'=(\Omega-1)\left[(3\gamma-2)\Omega+3\tilde{\Pi}\right],
\end{equation}
\begin{equation}
\label{eqn:Hprime}
H'=-H\left\{1+\frac{1}{2}\left[ (3\gamma-2)\Omega+3\tilde{\Pi}\right]\right\},
\end{equation}
and
\begin{equation}
\label{eqn:Piprimetis}
\tilde{\Pi}'=3\epsilon\gamma\Omega(\gamma-2)\left[1+\frac{H^{1-2s}\tilde{\Pi}}{(3\Omega)^{s}\alpha}\right]+\tilde{\Pi}\left[2+(3\gamma-2)\Omega+3\tilde{\Pi}\right],
\end{equation}
where $'prime'$ indicates a derivative with respect to the new time variable $\tilde{\tau}$. In the flat and expanding Universe $H$ is positive, so the dynamical equations (\ref{eqn:omega1}) - (\ref{eqn:Piprimetis}) are well defined. The present model consisting a single component, the non-relativistic viscous matter for which $\gamma=1,$ hence the density parameter will be $\Omega=1.$ Therefore, effectively the phase space can be described with dynamical equations (\ref{eqn:Hprime}) and (\ref{eqn:Piprimetis}). In the previous section, the behaviour of exact solutions for $s=1/2$ has analysed. But in this section, the phase space evolution is analysed for the choices $s=1/2$ and $s\neq1/2.$ 

\subsection{Choice 1: s=1/2}
For $s=1/2,$ the autonomous equations (\ref{eqn:Hprime}) and (\ref{eqn:Piprimetis}) become independent of each other, and hence the
dimension of the phase space will be reduced to one and (\ref{eqn:Piprimetis}) will then represents the evolution of this single dimensional phase space. Then, (\ref{eqn:Piprimetis}) can be expressed in terms of the variable $\omega=\tilde{\Pi}/\Omega$ as, 
\begin{equation}
\label{eqn:omega1TIS}
\omega'=f(\omega)=(\omega-\omega^+)(\omega-\omega^-),
\end{equation}
where we took $\Omega=1$ owing to the consideration of a single component and $\omega^+$ and $\omega^-$ are given as
\begin{equation}
\label{eqn:criticalpoints1DTIS}
\omega^{\pm}=\frac{\sqrt{3}\epsilon-3\alpha\pm\sqrt{(3\alpha-\sqrt{3}\epsilon)^2+36\alpha^2\epsilon}}{6\alpha}.
\end{equation}
The dynamical equation (\ref{eqn:omega1TIS}) gives the one dimensional phase plane evolution.
The equilibrium points are obtained by equating $\omega'=0,$ and are obviously $\omega^+$ and $\omega^-.$ From (\ref{eqn:omega1TIS}), we get $\omega^+>0$ and $\omega^-<0$ for all positive values of the model parameters $\alpha$ and $\epsilon.$ The equilibrium points (\ref{eqn:criticalpoints1DTIS}) can be read as $\omega^+\sim \frac{-3\alpha+\sqrt{9\alpha^2}}{6\alpha}=0$ and $\omega^-\sim \frac{-3\alpha-\sqrt{9\alpha^2}}{6\alpha}=-1,$ for a small value of $\epsilon.$ For the best estimated values of model parameters also we get $\omega^+=0$ and $\omega^-=-1.$ This shows that the critical points $\omega^+$ and $\omega^-$ represents the prior decelerated and a far future de Sitter expansion phases of the bulk viscous Universe respectively.
\begin{figure}
	\centering
	\includegraphics[scale=0.75]{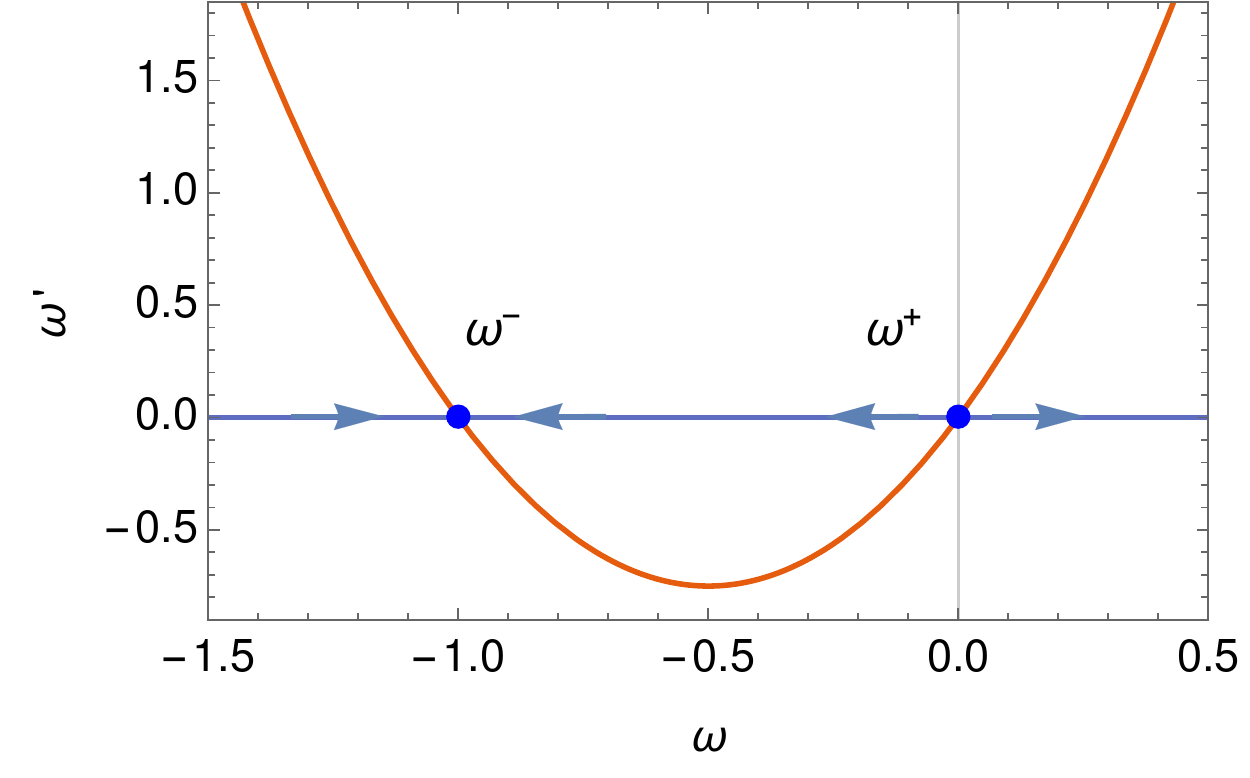}
	\caption{The evolution of phase space trajectory in the $\omega'\,-\,\omega$ space for the best estimated value of the model parameter, when $s=1/2.$}
	\label{fig:ppt}
\end{figure}

The phase space trajectory in the $\omega'-\omega$ plane will helps to study the evolution of the Universe without considering the exact analytic solutions. The phase space evolution according to  (\ref{eqn:omega1TIS}) is shown in figure \ref{fig:ppt}. The flow of vectorfield $\omega$ is on a line and its direction of evolution is determined from the stability of critical points \cite{Awad,CQG2}. A critical point $\omega_c$ becomes unstable, if $f'(\omega_c)>0,$ consequently a small perturbation around such a point will grow exponentially. On the other hand, if $f'(\omega_c)<0,$ then the critical point will be stable, then all small perturbations around that point will decline exponentially. Finally, if the slope $f'(\omega_c)$ alters its sign at the equilibrium point, then it will be a saddle point. The stability of critical points $\omega^\pm$ can be obtained as follows. From  (\ref{eqn:omega1TIS}) the expression for the slope is,
\begin{equation}
\label{eqn:omega2slope}
f'(\omega)=(\omega-\omega^+)+(\omega-\omega^-),
\end{equation}
where the condition $\omega^-<\omega<\omega^+$ is satisfied for the best estimated value of model parameters. At equilibrium points $\omega^{+}$ and $\omega^{-}$ the function $f'(\omega)$ become,
\begin{equation}
\label{eqn:omega2omegap}
f'(\omega^+)=(\omega^+-\omega^-)>0,
\end{equation}
\begin{equation}
\label{eqn:omega2omegam}
f'(\omega^-)=(\omega^--\omega^+)<0.
\end{equation}
From (\ref{eqn:omega2omegap}) and (\ref{eqn:omega2omegam}) it is evident that $\omega^+$ is an unstable equilibrium point and $\omega^-$ is a stable equilibrium point. In figure \ref{fig:ppt}, we can see that the flow of phase space trajectory is evolving from the unstable critical point $\omega^+$ and converges at the stable critical point $\omega^-.$ Therefore, the present truncated version of the viscous model is predicting the evolution of the Universe from an unstable decelerated to a stable accelerated de Sitter epoch in the late stages.

The exact solutions corresponding to the equilibrium points $\omega^\pm,$ can be obtained from  (\ref{eqn:Hprime}) as,
\begin{equation}
\label{eqn:HandaTISequlibriumpointsonebytwo}
H_{\omega^\pm}=\frac{2}{3}\frac{1}{(1+\omega^\pm)t},\qquad a_{\omega^\pm}=a_0t^{\frac{2}{3}\frac{1}{(1+\omega^\pm)}}.
\end{equation}
This clearly indicates that the case with $\frac{2}{3}\frac{1}{(1+\omega^+)}<1$ represents the decelerated behaviour while for $\frac{2}{3}\frac{1}{(1+\omega^-)}>1,$ the solution 
will represent an accelerating nature of the critical points $\omega^+$ and $\omega^-$ respectively. The properties of the equilibrium points are summarised in table \ref{Table:TISonebytwo}.
\begin{table}
	\centering
	\caption{\label{Table:TISonebytwo} Qualitative properties of the critical points $\omega^+$ and $\omega^-,$ when $s=1/2$}
	\begin{tabular}{|c|c|c|}
		\hline
		Critical points $\rightarrow$&$\omega^+$ & $\omega^-$  \\  [4pt]
		\hline
		$\omega$  & $0$& $-1$  \\ [4pt]
		\hline
		$q$& $0.5$ & $-1$ \\ [4pt]
		\hline
		Stability & Unstable  & Stable  \\ [4pt]
		\hline
	\end{tabular}
\end{table}
\subsection{Choice 2: $s\neq1/2$}
For deriving the exact solution, we choose $s=1/2,$ because for $s\neq1/2,$ it's difficult to get an analytical solution for the Hubble parameter. 
However, it is possible to have a qualitative analysis of the 
model is obtained for case $s \neq 1/2.$ For this, we modify 
the system of equations (\ref{eqn:Hprime}) and (\ref{eqn:Piprimetis}) as,
\begin{equation}
\label{eqn:hprimeneqs}
h'=\frac{3}{2}(2s-1)(1+\omega)h,
\end{equation}
\begin{equation}
\label{eqn:Omega1sneqonebytwo}
\omega'=3\left[\omega^2+\omega-\epsilon\left(1+\frac{h\omega}{3^s\alpha}\right)\right],
\end{equation}
where a new phase space variable is defined as $h=H^{1-2s}.$ Unlike in the previous case, for $s\neq1/2$ the phase space description is two dimensional with phase space variables $(h,\omega).$ Equating $h'=\omega'=0,$ in the above equations, we obtained two critical points as, 
\begin{equation}
P_1:\quad h=0, \qquad \omega=\frac{1}{2}(\sqrt{1+4\epsilon}-1)\sim 0,
\end{equation}
\begin{equation}
P_2:\quad h=3^s\, \alpha, \qquad \omega=-\frac{1}{2}(\sqrt{1+4\epsilon}+1)\sim-1.
\end{equation}

Since $h=H^{1-2s},$ for $s<1/2,$ the critical point $P_1$ represents a static Universe, owing to the zero value of the Hubble parameter. 
At the same time, $P_2$ represents a de Sitter epoch with a positive constant Hubble parameter. This implies that the option $s<1/2$ is not admissible with respect to the current observations regarding the evolutionary stages of the Universe.

Now, for the case $s>1/2,$ the fixed point $P_1$ will corresponds to the Hubble parameter $H\sim\infty$ and equation of state $\omega \sim 0$ for small value of $\epsilon.$ This means that $P_1$ represents a 
decelerating phase. For the same $s$ value, the point $P_2$ corresponds to a constant (positive) Hubble parameter and an equation of state $\omega \sim -1,$ which represents the late de Sitter phase. Therefore, for the option $s>1/2,$ this model can predict a transition from a decelerated epoch to the de Sitter phase during the late stages. However, the possibility of such a smooth transition actually depends on the nature of the stability of both $P_1$ and $P_2.$ 
\begin{table}
	\centering
	\caption{\label{Table:T2Sneqtis} Qualitative properties 
		of the critical points $P_1$ and $P_2,$ when $s\neq1/2$}
	
	\begin{tabular}{|c|c|c|}
		\hline
		Critical points $\rightarrow$&$P_1$ & $P_2$ \\ [4pt]
		\hline
		$h$  & $0$& $3^s \alpha$  \\ [4pt]
		\hline
		$\omega$& $0$ & $-1$ \\ [4pt]
		\hline
		$q$ & $0.5$  & $-1$ \\ [4pt]
		\hline
		Stability, $s>1/2$ & saddle & unstable \\ [4pt]
		\hline
	\end{tabular}
\end{table}

The stability nature of $P_1$ and $P_2$ can be obtained from the sign of eigenvalues of the corresponding Jacobian matrix. By linearising the system of equations (\ref{eqn:hprimeneqs}) and (\ref{eqn:Omega1sneqonebytwo}) about these critical points, we can obtain the Jacobian matrix $J(h,\omega)$ as,
\begin{equation}
\label{eqn:JacobianTIS}
J(h,\omega)=\left[
\begin{array}{cc}
\frac{3}{2} (2 s-1)h & \frac{3}{2} (2 s-1) (1+\omega ) \\
3 \left(1+2\omega-\frac{\epsilon h}{3^s \alpha }\right) & -\frac{\epsilon \omega }{3^{s-1}\alpha }
\end{array}
\right].
\end{equation}
Diagonalising $J(h,\omega),$ the eigenvalues $\lambda_1^\pm$ and $\lambda_2^\pm$ for $P_1$ and $P_2$ respectively are obtained as,
\begin{equation}
\label{eqn:eigenvalueP1TIS}
\lambda_1^{+}=3\sqrt{s-\frac{1}{2}}\quad \lambda_1^-=-3\sqrt{s-\frac{1}{2}},
\end{equation}
\begin{equation}
\label{eqn:eigenvalueP2TIS}
\lambda_2^+=\frac{3^{1-s}\epsilon}{\alpha} \quad \lambda_2^-=3^{1+s}\left(s-\frac{1}{2}\right)\alpha.
\end{equation}
When $s>1/2$, 
the eigenvalues, $\lambda_1^+>0$ and $\lambda_1^-<0$ implying that $P_1$ is a saddle point. On the other hand, both 
$\lambda_2^+$ and $\lambda_2^-$ are positive, hence $P_2$ 
is an unstable equilibrium point. The saddle character of $P_1,$ indicates that the Universe will continue to evolve from the decelerated epoch. But the unstable nature of $P_2$ is not a good sign, since it indicates the unstable end de Sitter epoch. These details are tabulated in table \ref{Table:T2Sneqtis}, where we have shown the values of the corresponding $q$ factor also.

The exact solution for $P_1$ when $s>1/2$ can be obtained by converting (\ref{eqn:hprimeneqs}) in terms of H, through a simple integration, we can get,
\begin{equation}
H=\frac{2}{3(1+\omega)\,t}
\end{equation}
In the limit $ t \to 0,$ the Hubble parameter $H \rightarrow \infty.$ 
Further integration of the above equation gives the scale factor as, $a\sim t^\frac{2}{3(1+\omega)}.$ The exact solutions at $P_2$ are given as,
\begin{equation}
\label{eqn:Handas>1/2TIS}
H=(3^s \alpha)^{\frac{1}{1-2s}}=\tilde{H}_0, \qquad a=e^{\tilde{H}_0 t}.
\end{equation}
These solutions are characterising the prior decelerated and de Sitter evolution of the bulk viscous Universe in the late time expansion respectively.

In summary, the phase analysis shows that for both $s>1/2$ and $s<1/2$ the evolution of the Universe predicted by the model is not compatible with the existing cosmological observation.  In the case of $s<1/2,$ the initial critical point corresponds to a static Universe and the later critical point gives an unstable de Sitter phase. For $s>1/2,$ even though the prior critical point represents a decelerated epoch, the second critical point implies again an unstable de Sitter epoch. These results imply that only the value $s=1/2$ is a feasible one, corresponding to which the model exhibits a realistic evolution of the Universe.

\section{Thermodynamic analysis}
\label{sec:ThermodynamicanalysisTIS}
\subsection{Generalised second law of thermodynamics}
The generalized second law (GSL) demands the change in sum of entropy of fluid components and that of horizon of the Universe will always increase with time \cite{Gibbons}. It can be expressed as,
\begin{equation}
\label{eqn:GSL}
S'_{m}+S'_{h}\geq0,
\end{equation}
where $ S_{m} $ is the matter entropy, $ S_{h} $ is the horizon entropy and $'prime'$ denotes a derivative with respect to a suitable cosmological variable. The horizon entropy is expressed as \cite{Davis},
\begin{equation}
\label{eqn:entropyhorizon}
S_h=\frac{A}{4l_P^2}k_B=\frac{\pi c^2}{l_P^2 H^2}k_B,
\end{equation}
where $A$ is the area of the Hubble horizon, $A=4\pi c^2/H^2,$ $l_P$ is the Planck length, $k_B$ is the Boltzmann constant, and $c$ is the velocity of light. The change in matter entropy can be obtained from Gibbs' relation,
\begin{equation}
T_m dS_m=c^2 V d\rho+(c^2 \rho +P_{eff})dV,
\end{equation}
and we consider derivative with respect to the scale factor,
\begin{equation}
T_m \frac{dS_m}{da}=c^2 V \frac{d\rho}{da}+(c^2 \rho+P_{eff})\frac{dV}{da}.
\end{equation}
The matter density is $\rho=\frac{3H^2}{8\pi G}$ and the volume can be expressed as $V=\frac{4\pi c^3}{3H^3}.$ We have the relation $\dot{\rho}+\frac{3H}{c^2}(c^2\rho+\Pi)=0,$ then $c^2\rho+\Pi=-\frac{\rho' c^2 a}{3}.$ Now, using these relations the Gibbs relation can be written as,
\begin{equation}
T_m S_m'=\frac{c^5 H'}{G H^2}\left(1+\frac{H'a}{H}\right)=-\frac{c^5 H'q}{G H^2},
\end{equation} 
where $G$ is the gravitational constant. In thermal equilibrium, the temperature of the horizon and that of the viscous matter are equal, $T_h=T_m.$ Then the variation in matter entropy becomes
\begin{equation}
\label{eqn:Sm'}
S_m'=-\frac{c^5 H' q}{G H^2 T_h}=-\frac{2\pi c^2 H' q}{l_P^2 H^3},
\end{equation}
where $T_h=\frac{H\hbar}{2\pi}$ is the Hawking temperature of the horizon in units of $k_B$. Using  (\ref{eqn:Sm'}) and the derivative of (\ref{eqn:entropyhorizon}), the GSL equation (\ref{eqn:GSL}) can be written as
\begin{equation}
\label{eqn:GSLgeneralTIS}
S'=S_h'+S_m'=-\frac{2\pi c^2}{l_P^2}\frac{H'}{H^3}(1+q)\geq0,
\end{equation}
\begin{figure}
	\centering
	\includegraphics[scale=0.6]{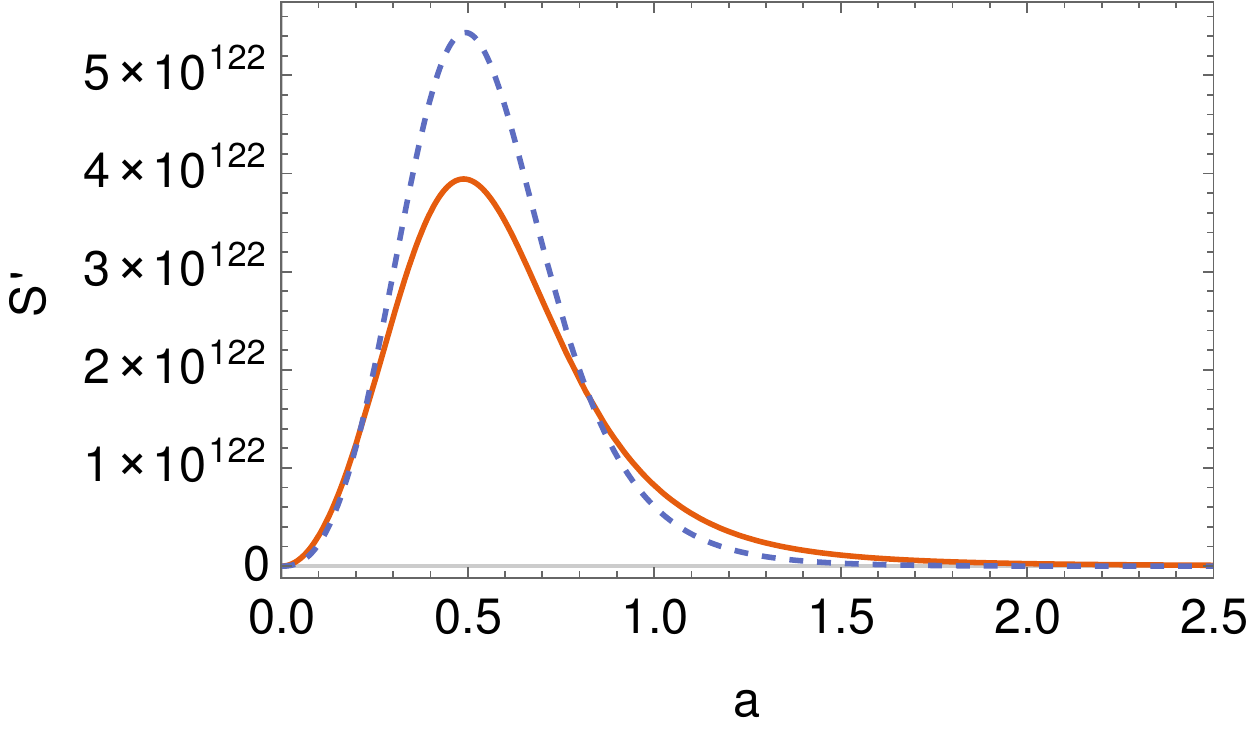}\,\includegraphics[scale=0.64]{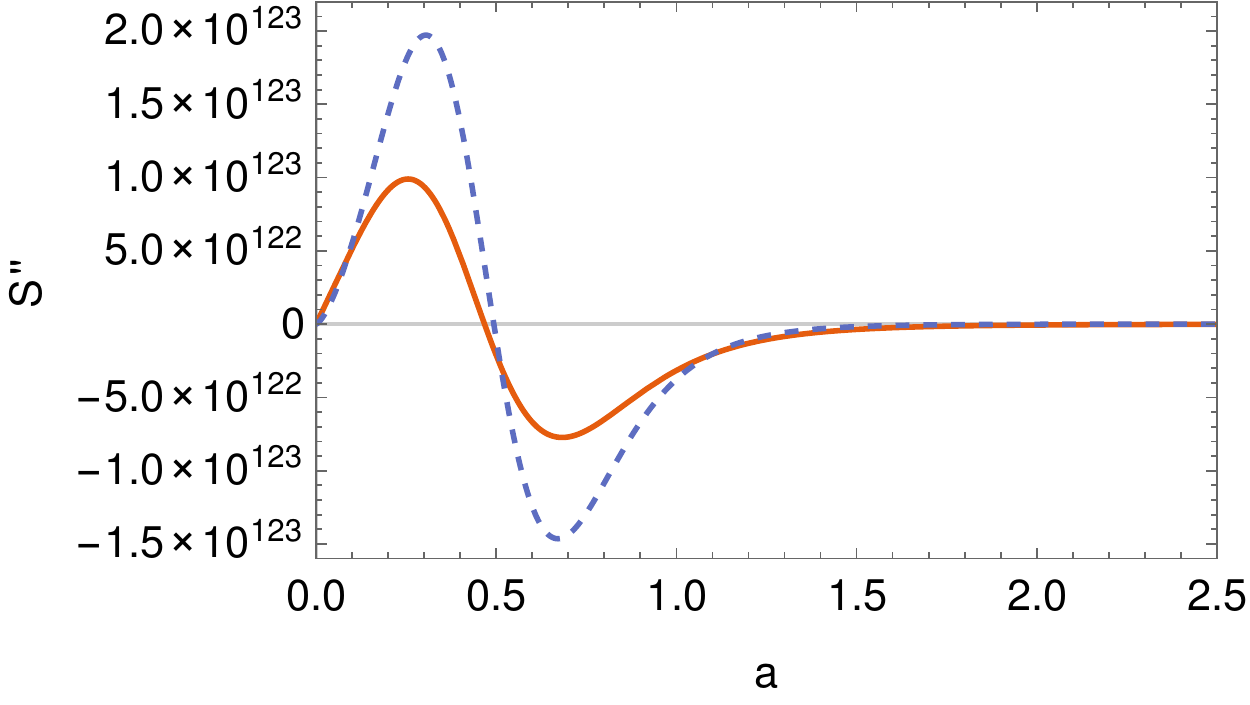}
	\caption{\label{plot:EntropyTIS}The evolution of $S'$ and $S''$ with scale factor for the best estimated values of the model parameters when $s=1/2$ for $\gamma=1$ (continuous line) and $\gamma=1.23$ (dashed line).}
\end{figure}
In the present model, when $s=1/2,$ we have $H\geq0,$ $H'\leq0,$ and $(1+q)\geq0,$ for the best estimated values of the parameters.
Then, from (\ref{eqn:GSLgeneralTIS}), we get $S'\geq0$ and hence the GSL is satisfied in this model. 
The evolution of $S'$ with scale factor for the best estimated value of parameters is plotted as shown in figure \ref{plot:EntropyTIS}. 
In figure \ref{plot:EntropyTIS}, the evolution of $S'$ first increases in the decelerated epoch, attains a maximum and then decreases in the recent accelerated phase of expansion. 
The maximum of $S'$ is representing the transition from decelerated to the current accelerated phase of expansion, a sudden variation in the slope of the curve is observed 
around transition redshift. The GSL is satisfied for both the choices $\gamma=1$ and $\gamma=1.23.$ The bulk viscous models based on the full IS theory is also 
satisfy the GSL \cite{EPJC1} throughout in the evolution of the Universe.

For $s>1/2,$ at the equilibrium point $P_1,$ the entropy derivative with respect to newly defined time ${\tau}$ can be expressed as,
\begin{equation}
\frac{dS}{d{\tilde{\tau}}}=\frac{9\pi c^2}{2 l_P^2}(1+\omega)^2 e^{3(1+\omega)\tilde{\tau}}.
\end{equation}
Thus we get $\frac{dS}{d{\tilde{\tau}}}>0$ at $P_1$ and $S'=0$ at $P_2$, ensures the fulfilment of GSL in the bulk viscous model for the choice $s>1/2.$

\subsection{Convexity condition of entropy}
In addition to GSL, an ordinary macroscopic system evolves to a stable thermodynamic equilibrium state must satisfy the convexity condition of entropy \cite{Pavon1}, which is given as
\begin{equation}
S''<0, \textrm{at least in the long run},
\end{equation} 
where $S=S_h+S_m.$
Taking again a derivative of $S'$ from (\ref{eqn:GSLgeneralTIS}) with respect to scale factor, 
we obtained
\begin{equation}
\label{eqn:S''generalTIS}
S''=-\frac{2\pi c^2}{l_P^2}\left[\frac{H'}{H^3}q'+(1+q)\left(\frac{H''}{H^3}-\frac{3H'^2}{H^4}\right)\right].
\end{equation}
The behaviour of $S''$ in the present model 
for $s=1/2$ is obtained by substituting the Hubble parameter from  (\ref{eqn:HTIS}) and 
its derivatives in (\ref{eqn:S''generalTIS}). The evolution of $S''$ for the best estimated values of the parameters is shown in figure \ref{plot:EntropyTIS}. 
Figure \ref{plot:EntropyTIS} explains that, $S''>0$ in the early decelerated phase, while in the later accelerated phase $S''<0$ 
and finally approach zero from below.
The change in sign of $S''$ occurred around transition time. The evolution of $S''$ corresponding to 
both $\gamma=1$ and $\gamma=1.23$ show similar behaviour. 
Hence, the present model admits bounded evolution of total entropy, and the model will achieve thermodynamic equilibrium 
through the maximisation of entropy. The possibility for 
any instabilities in the end stage is ruled out by the boundedness of entropy \cite{Callen1}. 
The full IS bulk viscous model exhibits a similar evolution of $S''$ \cite{CQG2}.

\section{Estimation of model parameters}
\label{sec:EstimationofmodelparametersTIS}
We have estimated the parameters, $\alpha$, $\epsilon$, $\tilde{\Pi_0},$ and $H_0,$ 
using observational data on type Ia Supernovae, Observational Hubble Data (OHD), Cosmic Microwave Background (CMB) shift parameter and acoustic peak parameter 
from Baryon Acoustic Oscillation (BAO) measurement. The parameter estimation is carried out for $\gamma=1$ and by treating $\gamma$ as a 
free parameter. We have adopted 
$\chi^2$ minimization method for extracting the parameters. 

For supernovae data, we have used the latest Pantheon sample, composed of 1048 
data points in the redshift range $0.01\leq z \leq 2.3$ \cite{Scolnic}.
The Pantheon sample is a compilation of 279 SNe Ia discovered by the Pan-STARRS1 medium deep survey, 
the distance estimates from the Sloan Digital Sky Survey (SDSS), Supernova Legacy Survey (SNLS) and from various low redshift and Hubble Space Telescope (HST) samples. A modified version of the Tripp formula \cite{Tripp} with two nuisance parameters calibrated to zero with the BEAMS with Bias Correction (BBC) method as proposed by Kessler and Scolnic is used for the distance estimator of the Pantheon sample \cite{Scolnic1}. Thus, the observational distance modulus can be written as,
\begin{equation}
\mu_{i}=m-M,
\end{equation}
where $m$ is the apparent magnitude and $M$ is a nuisance parameter, the absolute magnitude. The Pantheon sample give the corrected apparent magnitude for each Supernovae Ia. The luminosity distance $d_L$ in a flat Universe is expressed as
\begin{equation}
d_{L}(z,\alpha,\epsilon,\tilde{\Pi_{0}},H_{0},\gamma)=c(1+z)\int_{0}^{z} \frac{dz'}{H(z',\alpha,\epsilon,\tilde{\Pi_{0}},H_{0},\gamma)}.
\end{equation}
where $H(z',\alpha,\epsilon,\tilde{\Pi_{0}},H_{0},\gamma)$ is the Hubble parameter given in (\ref{eqn:HTIS}), and $c$ is the speed of light. The theoretical distance moduli $\mu_{th}$ and luminosity distance $d_L$ are related as
\begin{equation}
\label{distancemodulii}
\mu_{th}(z,\alpha,\epsilon,\tilde{\Pi_{0}},H_{0},\gamma)=5\log_{10}\left[\frac{d_{L}(z,\alpha,\epsilon,\tilde{\Pi_{0}},H_{0},\gamma)}{Mpc}\right]+25.
\end{equation}
The statistical $\chi^2$ function corresponding to the present model is written as
\begin{equation}
\chi_{SNIa}^{2}=
\sum_{i=1}^n \frac{\left[\mu_{th}(z,\alpha,\epsilon,\tilde{\Pi_{0}},H_{0},\gamma)-\mu_{i}\right]^2}{\sigma_{i}^2},
\end{equation}
where $\mu_{i}$ is the observational distance modulus obtained from the Pantheon sample, $n$ is the total number of data points, $\sigma_{i}^2$ is the variance of the $i^{th}$ measurement. The observational distance modulus $\mu_{i}$ is compared with $\mu_{th}$ calculated from (\ref{distancemodulii}) for different values of $z$.

\begin{figure}
	\centering
	\includegraphics[scale=0.5]{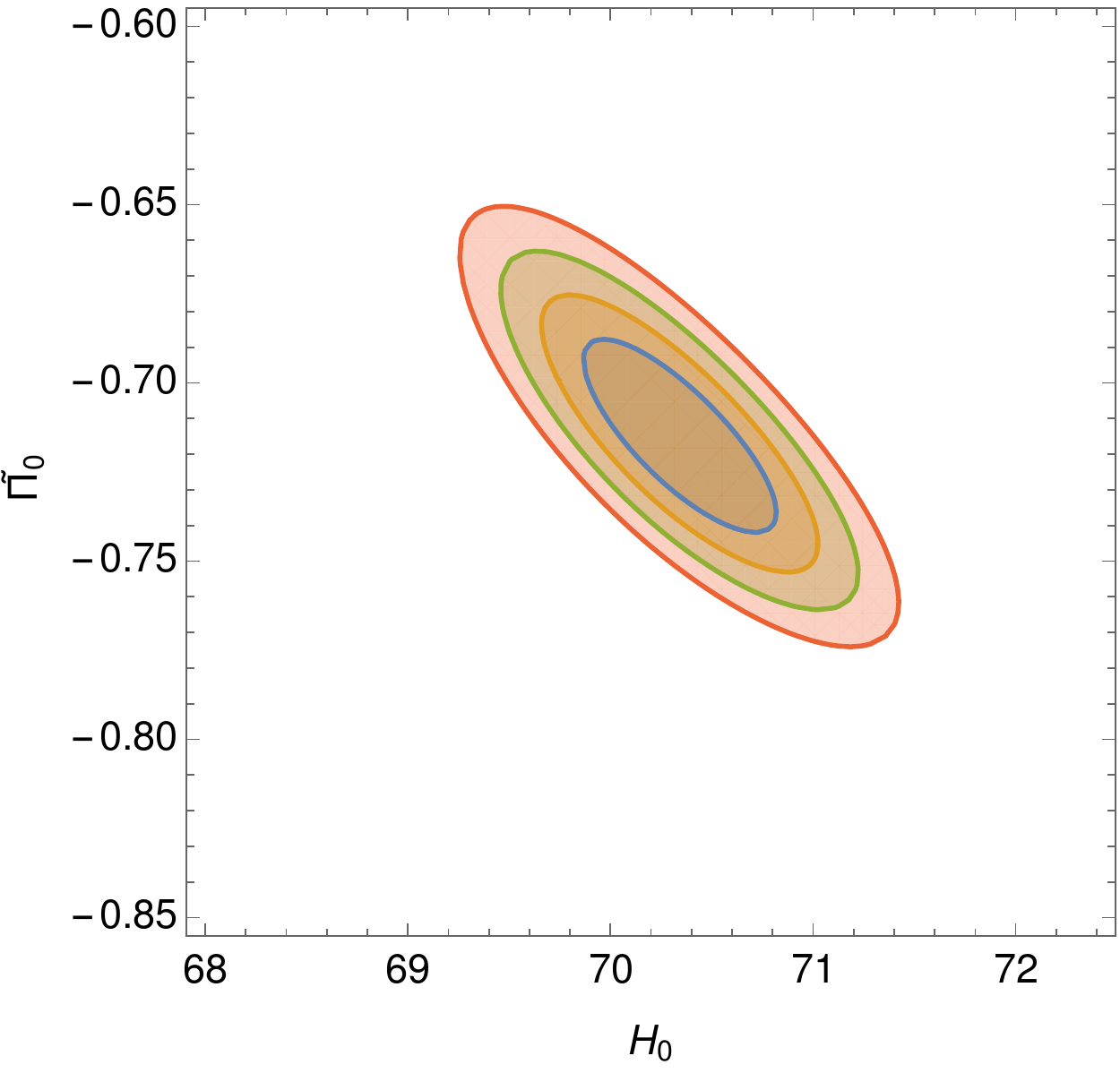}\,\includegraphics[scale=0.5]{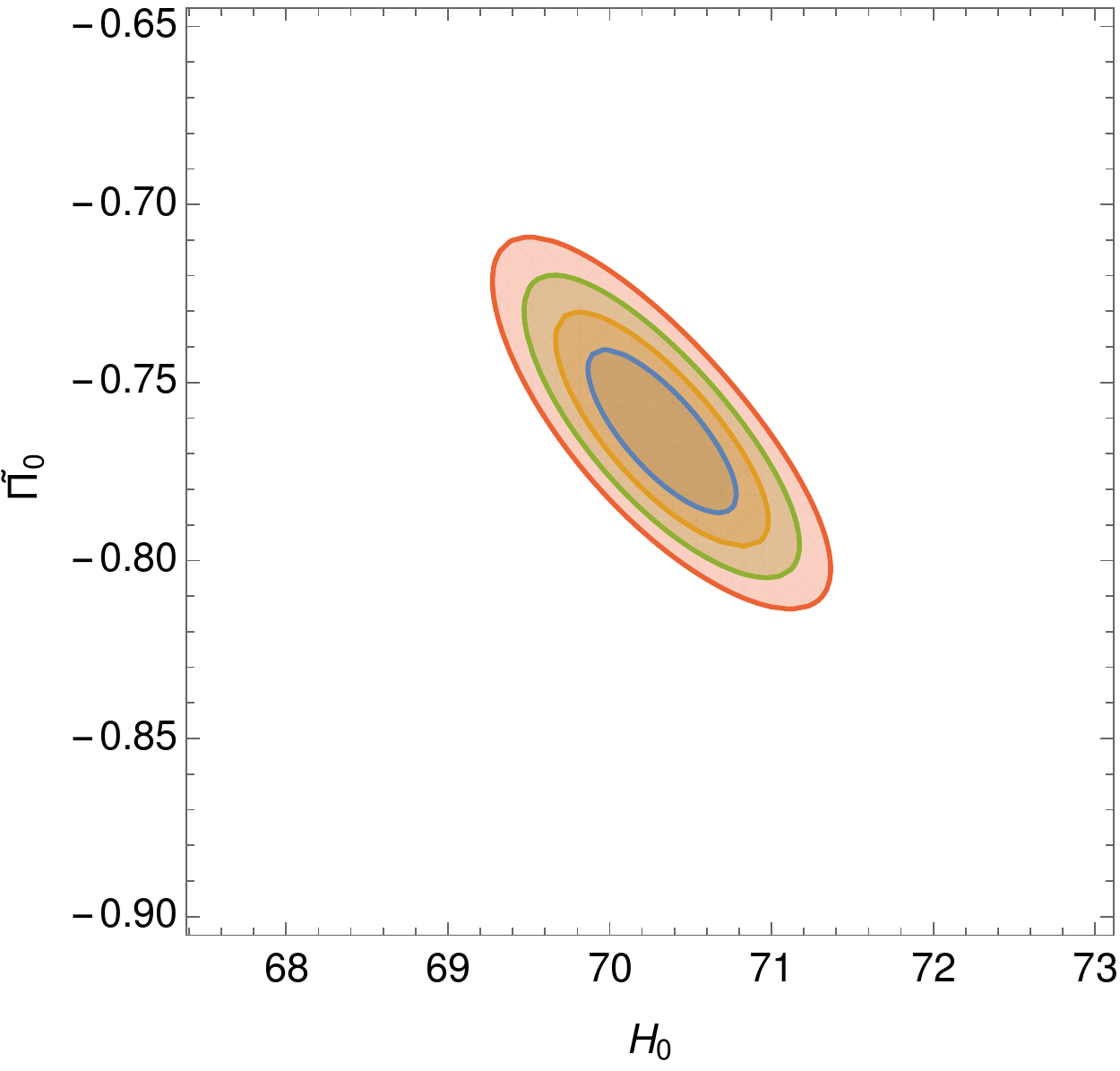}
	\caption{The confidence intervals  of the model parameters $H_0$ and $\tilde{\Pi_0}$ for $\gamma=1,$ left-hand panel and for $\gamma=1.23$ right hand panel, using the Pantheon sample. The contours corresponding to $68.3\%, 95.4\%,99.73\%$ and $99.99\%$ probabilities as one move from inside.}
	\label{fig:cpt}
\end{figure}

The best fit values of the model parameters are given in table \ref{Table:TISparameter}, where $\chi^{2}_{d.o.f.}$ is the $\chi^2$ function per degrees of freedom and $\chi^{2}_{d.o.f.}=\frac{\chi^{2}_{min}}{n-n'}$ where $n$ is the number of data points and $n'$ is the number of parameters in the model.
The confidence intervals for the model parameters $ H_0 $ and $\tilde{\Pi_{0}}$ are shown in figure \ref{fig:cpt}.

The Observational Hubble Data (OHD) determined using the cosmic chronometer method, consisting of $31$ $H(z)$ measurements in the redshift range $0.07<z<1.97,$ are used for constraining the model parameters \cite{Ryan}. Using the Hubble parameter (\ref{eqn:HTIS}),
the $\chi^2$ function can be expressed as,
\begin{equation}
\chi^2_{OHD}=\sum_{i=1}^n\frac{\left[H_{th}(z',\alpha,\epsilon,\tilde{\Pi_{0}},H_{0},\gamma)-H_i\right]^2}{\sigma_{i}^2,}
\end{equation} 
where $H_{th}(z',\alpha,\epsilon,\tilde{\Pi_{0}},H_{0},\gamma)$ is the Hubble parameter in this model and $H_i$ is from measurements of OHD and $\sigma_{i}$ is variance in each measurement.

The CMB shift parameter $\mathcal{R}$ is given by the relation,
\begin{equation}
\mathcal{R}=\sqrt{\Omega_{m}}\int_{0}^{z_{ls}}\frac{d z}{h(z),}
\end{equation}
where $\Omega_m$ is the density parameter and $h(z)=H/H_0$ and $z_{ls}$ is the redshift at the surface of the last scattering. The Planck 2018 observations predicts $\mathcal{R}_{obs}=1.7502\pm 0.0046$ and $z_{ls}=1089.92$ \cite{Chen}. Now the $\chi^2$ function of this model can be written as
\begin{equation}
\chi^2_{CMB}=\frac{\left[\mathcal{R}_{th}(\alpha,\epsilon,\tilde{\Pi_{0}},\gamma)-\mathcal{R}_{obs}\right]^2}{\sigma_{i}^2},
\end{equation}
where $\mathcal{R}_{th}(\alpha,\epsilon,\tilde{\Pi_{0}},\gamma)$ is the theoretical shift parameter calculated from this viscous model and $\sigma_{i}$ is the variance of the measurement $\mathcal{R}_{obs}.$ Since this model assumes the viscous matter dominated Universe, then we have taken $\Omega_m\sim  1.$ 

The baryon acoustic oscillation (BAO) peak parameter $\mathcal{A}$ is defined as,
\begin{equation}
\label{eqn:BAOpeak}
\mathcal{A}=\frac{\sqrt{\Omega_m}}{{h(z_{ap})}^{1/3}}\left( \frac{1}{z_{ap}}\int_{0}^{z_{ap}}\frac{dz}{h(z)}\right)^{2/3},
\end{equation}
where $z_{ap}$ is the redshift of the acoustic peak parameter. The observed value of SDSS-BAO peak parameter is $\mathcal{A}=0.484\pm 0.016$ at the redshift $z_{ap}=0.35$ \cite{BlakeBAO}. The $\chi^2_{BAO}$ takes the form
\begin{equation}
\chi^2_{BAO}=\frac{\left[\mathcal{A}_{th}(\alpha,\epsilon,\tilde{\Pi_{0}},\gamma)-\mathcal{A}_{obs}\right]^2}{\sigma_{i}^2},
\end{equation}
where $\mathcal{A}_{th}(\alpha,\epsilon,\tilde{\Pi_{0}},\gamma)$ is the theoretical acoustic peak parameter obtained in this model using (\ref{eqn:BAOpeak}) and $\sigma_{i}$ is the variance of measured $\mathcal{A}_{obs}.$

The model parameters are estimated first using the Pantheon sample and then the combinations, Pantheon+OHD, Pantheon+OHD+CMB and Pantheon+OHD+CMB+BAO. 
The constrained values of the parameters are given in table \ref{Table:TISparameter}. The Pantheon data provides a good fit to the truncated viscous model of the Universe. 
The combination, Pantheon+OHD, and Pantheon+OHD+CMB provides goodness of fit with $\chi^2_{d.o.f.}$ around one. The best fit values of model parameters obtained from
the data combination, Pantheon+OHD+CMB gives $\mathcal{R}\sim 1.76$ corresponding to $\gamma=1$ and $\mathcal{R}\sim 1.75$ corresponding to $\gamma=1.47,$ indicates the adaptability of 
the truncated viscous model in explaining the early phase observational data. However, the $\chi^2$ minimum for the
combination of Pantheon sample+OHD+CMB+BAO gives a relatively high value, indicates an inconsistency in the fitting of the BAO data in the truncated viscous model. For $\gamma=1$ the value of $\epsilon$ is of the order of 
$10^{-8}$ for both Pantheon and Pantheon+OHD. But for the other data sets its value is relatively too high. While treating $\gamma$ as a free parameter, the $\epsilon$ value is 
around $10^{-8}$ for all the data combinations. It should also be noted that for both cases, the Hubble constant values are in concordance with the latest observation corresponding to the data sets, Pantheon and Pantheon+OHD.
\begin{table}
	\centering
	\caption{\label{Table:TISparameter} The best estimated values of 
		the model parameters and the $\chi^2$ minimum values 
		corresponding to $\gamma=1$ and $\gamma\neq 1$ respectively. 
		The $1\sigma$ $(68.3\%)$ uncertainties of the confidence 
		level of $H_0$ and $\tilde{\Pi_{0}}$ are given.}
	\begin{tabular}{|c|c|c|c|c|}
		\hline 
		Parameters & Pantheon &Pantheon+ & Pantheon+ &Pantheon+ \\
		{}&{}&OHD&OHD+CMB&OHD+CMB+BAO \\ [10pt]
		\hline
		$\alpha$& 1.00&1.00&0.45&0.30\\  [4pt]   \hline
		$\epsilon$ & $5.42\times 10^{-8}$ & $9.34\times 10^{-8}$ & $0.47$ & $0.39$ \\ [4pt]  \hline             
		$\tilde{\Pi_0}$ & $-0.72^{+0.03}_{-0.02}$&$-0.71^{+0.03}_{-0.03}$&$-0.77^{+0.01}_{-0.01}$&$-0.64^{+0.01}_{-0.01}$\\ [4pt]  \hline
		$ H_{0} $ & $70.34^{+0.46}_{-0.39}$& $69.69^{+0.52}_{-0.48}$&$66.97^{+0.24}_{-0.26}$&$65.18^{+0.28}_{-0.25}$ \\ [4pt] \hline
		$\gamma$&$1$&$1$&$1$&$1$\\ [4pt]  \hline
		M&$-19.35$&$-19.37$&$-19.47$&$-19.49$\\ [4pt] \hline
		$ \chi^{2}_{min} $ & $1035.68$&$1050.55$&$1058.60$&$1682.90$  \\[4pt]  \hline
		$ \chi^{2}_{d.o.f.} $ & $0.99$ &$0.98$&$0.98$&$1.56$\\ [4pt] \hline\hline 
		$\alpha$& $1.00$&$0.99$&$0.99$&$1.01$\\[4pt]  \hline
		$\epsilon$ & $9.93\times 10^{-8}$ & $9.87\times 10^{-8}$ & $5.51\times 10^{-8}$ & $5.43\times10^{-8}$ \\ [4pt]  \hline        
		$\tilde{\Pi_0}$ & $-0.76^{+0.02}_{-0.02}$&$-0.74^{+0.03}_{-0.02}$&$-0.81^{+0.01}_{-0.01}$&$-0.73^{+0.01}_{-0.01}$\\[4pt]  \hline
		$ H_{0} $ & $70.32^{+0.48}_{-0.52}$& $69.05^{+0.35}_{-0.45}$&$66.41^{+0.29}_{-0.31}$&$65.00^{+0.20}_{-0.20}$ \\[4pt]  \hline
		$\gamma$&$1.23$&$1.14$&$1.47$&$1.37$\\[4pt]  \hline
		M&$-19.36$&$-19.39$&$-19.49$&$-19.50$\\ [4pt] \hline
		$ \chi^{2}_{min} $ & $1032.73$&$1048.05$&$1064.62$&$1728.08$ \\[4pt]  \hline
		$ \chi^{2}_{d.o.f.} $ & $0.99$ &$0.97$&$0.99$&$1.60$\\[4pt]  \hline
	\end{tabular}
\end{table}

For the best fit values of the model parameters, the dimensionless value of the viscous coefficient is 
$\alpha \sim 1$ using the Pantheon Supernovae Ia sample. The corresponding dimensionful quantity is around $\bar \alpha = \frac{c^2 H_0}{24\pi G \alpha}  \sim 4.08\times10^{7}$Pa s. This value is roughly in agreement with the value obtained by Velten and Schwarz \cite{Velten} and also by Sasidharan and Mathew \cite{Athira} 
using the Eckart model.
In \cite{Brevikrev1,BrevikVisEst2} Brevik et al. have extracted 
a range, $10^4 \textrm{Pa s} <\bar\alpha <10^7 \textrm{Pa s}$. 
The predicted range is many orders of magnitude higher than the viscosity of normal water at atmospheric pressure and room temperature. However, the
Universe is such a vast system, and the viscosity associated with the dark matter is not so comparable with ordinary macroscopic systems and hence 
such a relatively high value can not be ruled out.

\section{Conclusions}
\label{sec:ConclusionsTIS}
We analysed the evolution of the late Universe with a viscous dissipative mechanism using the truncated version of the Israel-Stewart theory of relativistic bulk viscosity formalism. The causal IS theories, the full IS and its truncated version, introduce a finite relaxation time $\tau$ as an additional parameter compared to the non-causal Eckart theory, where the relaxation time is taken as zero. A non-zero positive value of $\tau$ is essential for ensuring the causality condition. Among the causal theories, the full IS theory reduces to the truncated IS formalism (\ref{eqn:TIS}) under the condition $|\Pi|\ll\rho.$ However this does not alter the causality and stability conditions in the truncated theory. Hence, the truncated IS approach can be considered as an independent description of the bulk viscous pressure evolution. Some earlier authors have assumed such an independent status for these theories for describing the early inflation of the Universe in dissipative models. The non-negative entropy production in the truncated version also supports its independent nature.

There exists different opinions in the literature regarding the possibility of the inflationary solution in the dissipative models. Moreover, the possibility depends on whether one uses the full IS theory or the truncated version. Our main motivation is to search whether there exists any such dependence in extracting the possibilities of late acceleration in the dissipative models, on the nature of the dissipative theories like the truncated IS model or the full version of the IS theory or even the non-causal Eckart formalism. Earlier analyses
of the bulk viscous models in the literature have shown that it is possible to obtain solutions to explain the late accelerating phase of the Universe using both the Eckart formalism and the full IS theory. In the truncated version, we have derived analytical solutions for the late accelerating phase of the Universe.

Even though these three formalisms could predict the late accelerating epoch driven by the negative pressure generated by the bulk viscosity, 
there exist strong differences between them. Our analysis has shown that the truncated model may appear to be more favoured by cosmological observations.

We have derived the Hubble parameter (\ref{eqn:HTIS}) of the Universe obeying the truncated IS equation by assuming bulk viscosity and relaxation time of the form $\xi=\alpha\rho^s$ and $\tau=\frac{\alpha}{\epsilon \gamma(2-\gamma)}\rho^{s-1}$ respectively with the choice $s=1/2.$ Under the asymptotic limits the Hubble parameter behaves as $H\sim a^{-3/2}$ as $a \to 0,$ which corresponds to the early decelerating phase and $H\sim \,\textrm{constant} $ as $a \to \infty,$ corresponding to the 
pure end de Sitter epoch. The expansion profiles of the deceleration parameter and the equation of state parameter are found to be satisfying the respective asymptotic behaviours, $q \to 0.5, \omega \to 0$ as $a \to 0$ and $q \to -1, \omega \to -1$ as $a \to \infty.$ The implied transition from the prior decelerating epoch to the late acceleration is found to occurs at redshift around $z_T\sim 0.73.$ 
So, the background expansion history of the Universe in this truncated model is compatible with the current observation and predicts a pure de Sitter epoch as the end phase like the standard $\Lambda$CDM. However, we are more interested in comparing the results with the corresponding scenario with the full IS model and the Eckart theory. 

Let us first compare the background evolution of the cosmological parameters in the truncated model with the other two models.
Earlier studies on the full IS theory \cite{CQG2} have revealed that the
Hubble parameter assumes the form $H\sim a^{-2.8}$ as $a\rightarrow0,$ 
while studies on the Eckart viscous model indicate that $H\sim a^{-3.4}$ as $a \to 0$ \cite{Athira}. So in both these cases, the Hubble parameter decreases comparatively slower than in the case of ordinary matter dominated case. Consequently, the state of the cosmic component might be different from that of ordinary dark matter. In the same references, the equation of state of the viscous matter can be calculated for the same condition as, $\omega\to 0.88$ with a corresponding deceleration factor, $q\to 1.83$ for the full IS model, and in the Eckart viscous model, the parameters obtained as $\omega\to 1.3$ with $q\to 2.4.$ So the viscous matter in the prior decelerated epoch in both the full IS model, and the Eckart model has stiff nature.
Compared to these two, as per our analysis, the truncated model predicts a prior decelerated epoch with the equation of state $\omega=0$ corresponding to a deceleration parameter of $q=0.5.$ Let us now contrast the truncated model with the other models corresponding to the future asymptotic limit as $a\rightarrow\infty.$ In this limit, the full causal model predicts a quintessence behaviour with $H \sim a^{-0.2},$ $ q \to -0.82$ and $\omega\to -0.88$ i.e., an ever decreasing nature for the Hubble parameter, while $H \sim \,\textrm{constant}$ in the Eckart model, which represents the de Sitter evolution. 
In the truncated model, the evolution corresponds to a pure de Sitter epoch with $q\to -1,$ $\omega\to-1.$ So compared to both the Eckart model and the full IS model the truncated model predicts a similar evolution of the Universe as that in the case of the standard $\Lambda$CDM model. But unlike in the $\Lambda$CDM, we don't need any fictitious dark energy; instead, the acceleration is now generated by the more physical viscosity associated with the matter sector.

We have constrained the model parameters using the Pantheon sample and the combinations of Pantheon+OHD, Pantheon+OHD+CMB, and Pantheon+OHD+CMB+BAO.  Our study pointed out good adaptability of the truncated IS model in explaining the late accelerated expansion of the Universe. The analysis shows that Pantheon data provides a good fit for the truncated viscous model of the Universe. The combined analysis of the Pantheon+OHD and Pantheon+OHD+CMB yields goodness of fit of the model with $\chi^2_{d.o.f.}$ around one. However, the $\chi^2$ minimum for 
the	combination of Pantheon sample+OHD+CMB+BAO gives relatively high value. The magnitude of the bulk viscosity is obtained as $\bar{\alpha}\sim 4.08\times10^7$Pa s for $\gamma=1.$ The estimated value of the viscosity is consistent with its predicted range. Nevertheless, this value is high compared to the viscosity of an ordinary  macroscopic system. A major reason for the reasonable behaviour of the truncated model is lying in the value of the parameter $\epsilon,$ the causality parameter. This parameter determines the speed perturbations as $c_b^2=\epsilon(2-\gamma).$ For $\gamma=1$ we have $c_b^2=\epsilon,$ i.e., the causality parameter $\epsilon$ is account for $c_b^2.$ We 
extracted this parameter 
using the supernovae type Ia data as
$\epsilon\sim5.42\times 10^{-8}.$ In contrary, the 
best fit value in full IS viscous model is $\epsilon\sim0.39$ for $\gamma=1$ \cite{CQG2}. 
The extremely small value that we have extracted for $\epsilon$ is coinciding with the upper limit of the range of this parameter 
$10^{-11}\ll  \epsilon (= c_b^2)\leq 10^{-8}$ obtained in ref. \cite{Piattella}. 
Another plus point is the age prediction in the truncated model, a difficulty related to the prediction of the age of the Universe come across in the Eckart and the full IS viscous models are successfully sorted out in the present truncated viscous model. In the present model, for $\gamma=1,$ the deduced age of the Universe is around $13.66\, \,\textrm{Gyr}$ for the best estimated values of parameters. Obviously, the predicted age is
consistent with the recent observations {\cite{AghaninPlanck2018}}.

The dynamical system analysis of the truncated model, for the choice $s=1/2,$ elucidates that the future de Sitter phase will be a stable equilibrium point while the early decelerated phase is an unstable equilibrium. So, we could affirm that the Universe is evolving towards a stable equilibrium state. We further extended the phase space analysis for the choice of $s\neq1/2.$ When $s<1/2$ the solutions are failing to explain the standard evolution of the Universe, and hence we have ruled out this choice in explaining the model. While $s>1/2$, the early decelerated phase is saddle point and the future de Sitter epoch will be an unstable equilibrium point. Therefore an evolution towards stable equilibrium could not be achieved in this choice. Even though the full IS model, with $s=1/2,$ gives a stable end epoch, but it is of quintessence nature, and with $s>1/2,$ it also yields an unstable 
de Sitter equilibrium endpoint. So that $s=1/2$ is a favourable choice for the description of viscosity evolution in full and truncated dissipative models.

Thermodynamic analysis of the model shows that, for $s=1/2,$ the generalised second law of thermodynamics is satisfied throughout in the evolution of the Universe. The convexity condition of entropy, $S''<0,$ satisfied in the long run of expansion of the Universe, indicates the expansion towards a state of maximum entropy as in the evolution of an ordinary macroscopic system. Hence, the model admits bounded evolution of total entropy and achieves thermodynamic equilibrium in the end de Sitter epoch. 

The exact status of any model is to be confirmed with the studies on perturbation growth. But at this juncture, we can add some comments in this regard. 
Acquaviva et. al. \cite{Acquaviva}, have 
analysed the evolution of density contrast in 
viscous models 
for both $s>1/2$ and $s<1/2$ and compared with the $\Lambda$CDM model. 
For $s<1/2,$ the perturbation growth in the truncated version shows a substantial deviation from that in the standard $\Lambda$CDM model only during the late time evolution of the Universe. But for $s>1/2,$ the deviation in the growth of perturbation in the truncated model starts to occur from early time itself \cite{Acquaviva}. Therefore one may expect that, for $s=1/2,$ there can be a difference in the growth of perturbation in the truncated model 
compared to the standard $\Lambda$CDM model that starts from the early stage and persists till the late time evolution of the Universe.

In conclusion, our investigation on the truncated IS viscous model, for the choice $s=1/2$ is consistent with the cosmological observations, it predicts a prior decelerated epoch $q=0.5,$ $\omega=0$ and 
which further evolves to a stable pure de Sitter phase. Also, prediction of the age of the Universe from this model is compatible with the observational values than that of the other viscous models, viz. the Eckart and the full IS model.

\section*{Acknowledgments}
We are thankful to the referees for the useful comments, which helped to improve the manuscript. One of the authors TKM is thankful to IUCAA, Pune for the hospitality during the visits. JMND is thankful to UGC, Government of India for the support through the BSR fellowship.


\end{document}